# Towards decoding the structure of water-magnetite interface

Salvatore Romano[1,2,*], Harsharan Kaur[3,*], Pablo Montero de Hijes[1], Moritz Zelenka[3,6], Moritz Eder[4], Min-Ju Choi[8], Yingge Du[8], Gareth S. Parkinson[4], Yair Litman[7], Ellen H. G. Backus*[3,6], and Christoph Dellago†[1,5]

[1]Faculty of Physics, University of Vienna, Boltzmanngasse 5, 1090 Vienna, Austria
[2]Vienna Doctoral School in Physics (VDSP), University of Vienna, Boltzmanngasse 5, 1090 Vienna, Austria
[3]Institute of Physical Chemistry, Faculty of Chemistry, University of Vienna, Währinger Straße 42, 1090 Vienna, Austria
[4]Institute of Applied Physics, TU Wien, Wiedner Hauptstraße 8–10/134, 1040 Vienna, Austria
[5]Research Platform on Accelerating Photoreaction Discovery (ViRAPID), University of Vienna, 1090 Vienna, Austria
[6]Vienna Doctoral School in Chemistry (DoSChem), University of Vienna, Währinger Straße 42, 1090 Vienna, Austria
[7]Max Planck Institute for Polymer Research, Mainz, Germany
[8]Physical Sciences Division, Pacific Northwest National Laboratory, Richland, WA 99352, USA
[*]These authors contributed equally to this work.

**Abstract**

The molecular structure of water at iron-oxide interfaces is determined by the protonation state, orientation and population of interfacial OH species. These microscopic features can be decoded by combining surface-specific vibrational spectroscopy with reactive and accurate molecular dynamics simulations. In this way, it could be possible to connect the measured nonlinear optical response to specific molecular configurations at the magnetite–water interface, resolving the interfacial water structure.

We investigated magnetite films in contact with liquid water at neutral and alkaline pH using phase-resolved sum-frequency generation (SFG) spectroscopy in the OH/OD stretching region and Fresnel correction of the optical response. In parallel, neural-network-potential molecular dynamics simulations were used to model the water-magnetite interface and identify interfacial water species, quantify their orientations, hydrogen-bonding environments and lifetimes, and compute species resolved SFG spectra.

The combined analysis assigns distinct spectral features to dissociated surface OH groups, hydroxyls and water molecules adsorbed on octahedral iron sites, water bound at dissociation sites and water molecules above tetrahedral rows. The simulations reveal a reactive interfacial hydroxyl network with nanosecond-scale rearrangements and strong species-dependent hydrogen bonding. Experimentally, the SFG spectra show pronounced pH dependence, consistent with changes in surface charge and hydroxylation. A systematic difference between measured and computed SFG spectra indicates that the assumed surface model of magnetite does not fully represent the aqueous interface. This discrepancy provides a molecular constraint for improved models of magnetite–water interfaces and highlights SFG spectroscopy/ML-MD comparison as a valuable route to resolving oxide–water interfacial structures.

## 1 Introduction

Iron oxides are earth-abundant minerals with various crystallographic phases and oxidation states [1–4]. Among them, magnetite ($Fe_3O_4$) plays a key role across a wide range of fields, including geology and weathering [1, 5], electrochemistry [6, 7], atmospheric chemistry [8], catalytic reactions [9, 10], solar water splitting [2, 9], and biomedical [11]. Since $Fe_3O_4$ frequently exists in contact with aqueous solutions, the water-magnetite interface has been studied extensively, particularly for the stable (001) and (111) surfaces of $Fe_3O_4$ on single-crystals or as thin-films [2, 3, 10, 12–17]. These studies have significantly advanced our understanding of the thermodynamic stability, adsorption, dissolution, and passivation tendency of the $Fe_3O_4$ surfaces in contact with water molecules, both through experimental investigations under ultrahigh vacuum (UHV) conditions and computational approaches using Density Functional Theory (DFT) calculations [18–20].

To date, most studies of water-magnetite interactions have been conducted under UHV conditions, where an inter-connected assembly of partially dissociated water molecules forms [21] along with pre-adsorbed surface hydroxyls [22], and an ordered iron oxyhydroxide phase is found [23] on the $Fe_3O_4$ (001) surface. However, the behavior of the magnetite surface in contact with a macroscopic amount of water has not been explored experimentally. Computationally, studies are limited to classical simulations or short DFT-based simulations [24–26]. In a previous study [27] we developed a Neural Network Potential (NNP) of Behler-Parrinello architecture [28–30] for the water-magnetite interface with the magnetite exposing the Subsurface Cation Vacancy (SCV) reconstructed surface [31]. Simulations performed with this NNP provided key insights into the dissociation mechanism and the diffusion of water on the surface.

In this work, we combine Sum Frequency Generation (SFG) spectroscopy [32–34] with NNP-based molecular dynamics (MD) simulations to resolve the molecular structure of water at the magnetite–water interface. The simulations enable the identification of the OH species present at the surface, including dissociated OH at the surface interacting with hydrogen bonded water molecules on top of them, and water and OH group adsorbed on octahedrally coordinated irons. We provide their orientations, hydrogen-bonding environments, and dynamical stability. By directly computing the SFG response, we establish a molecular-level assignment of the experimental spectral features to specific interfacial OH species. While this combined approach yields a consistent interpretation of the spectroscopically active species, a systematic discrepancy in the net orientation of interfacial water molecules emerges between the measured and computed SFG spectra. Rather than indicating a failure of either method, this divergence points to limitations of the currently accepted structural model of the magnetite surface in contact with liquid water. As such, the comparison between experiment and simulation provides critical constraints that motivate alternative interfacial models and offer new insight into the structure, reactivity, and protonation state of the magnetite under aqueous conditions.

## 2 Methods

### Vibrational sum frequency generation (VSFG)

A femtosecond Ti:sapphire regenerative amplifier (Astrella, Coherent) generates 800 nm pulses with a duration of 40 fs and ∼7 mJ of pulse energy at a repetition rate of 1 kHz. 2 mJ laser output is used to pump an optical parametric amplification (TOPAS PRIME, Light Conversion) combined with a non-colinear difference frequency generation (NDFG) stage (Light Conversion) to yield an IR beam with FWHM of ∼300 $cm^{-1}$. Another 1 mJ of the 800 nm laser beam is guided to a home-built pulse shaper to narrow the pulse bandwidth to ∼20 $cm^{-1}$ and a power of 10 $\mu$J. The phase-resolved SFG set-up was constructed similar to that reported in literature [35]. Both the visible and IR beams are focused using a fused silica (LA4716-B, Thorlabs) and $CaF_2$ lens (LA5042, Thorlabs) respectively, on the local oscillator (LO) generator, which is a 150 nm sputter-deposited ZnO layer on a $CaF_2$ substrate (2 mm x 25 mm, Crystal GmbH). The resulting SFG and the visible and IR beams are collimated using an off-axis parabolic mirror (60°, MPD246-P01, Thorlabs) and re-focused onto the sample with a 60° off-axis parabolic mirror (MPD246-P01, Thorlabs). Before re-focusing, the LO-SFG beam undergoes a time-delay by passing through a 1 mm thick glass plate placed after the collimating parabolic mirror. In order to obtain a conventional SFG spectrum, the LO-SFG signal is blocked before the sample. The visible and IR beams are inclined on the sample with respective incident angles of ∼60° and ∼40° relative to the surface normal. The reflected SFG sig-nal from sample and LO-SFG signal are collimated and focused on the spectrometer (Shamrock 303i, 1200 g/mm, Andor) equipped with an electron-multiplied charge-coupled device (EMCCD, Newton) camera. All measurements were performed in SSP polarization mode, which is a combination of S-polarized: SFG beam, S-polarized: visible beam, and P-polarized: IR beam. The laser setup was purged with dry air to limit the absorption of the infrared beam by water and carbon dioxide in the atmosphere. All measurements were performed under ambient pressure in a temperature controlled lab at 20°C. We refer to the supporting information (SI) for additional details on the measurement cell, solution preparation and equilibration, calculations on the phase of the $Fe_3O_4$ sample, the overall phase uncertainty and the SFG data analysis.

### Computational details

Technical details about the magnetite/water NNP used in this study, i.e. its development, accuracy and range of applicability, can be found in Ref. [27]. All our NNP-based molecular dynamics (MD) simulations have been run with the software package n2p2 [36–38] interfaced with LAMMPS [39]. The simulation setup consists of a magnetite slab composed of 1152 oxygen (O) and 800 iron (Fe) in contact with a column of liquid water 55 Å with 2176 water molecules (started in [27]). The two SCV interfaces present (due to the periodic boundary conditions (PBCs) applied in all directions) are treated as two independent simulations with 26 Å of water. For the statistical analysis, we ran 29 ns saved every 0.5 ps and for the calculation of the spectra we ran four independent trajectories of 1.3 ns each saved every 0.04 ps. In both cases, the temperature was held constant at 300 K with a Langevin thermostat applied only in the interior of the magnetite slab. The volume was fixed and the pressure fluctuated around zero. The used timestep was 0.5 fs.

By construction, Neural Network Potential (NNP) are reactive. In simulations, atoms are considered individually without imposing any predefined molecular connectivity. Therefore, additional analysis is required to determine the molecular structure of the system. To identify water molecules and OH groups, each hydrogen atom is assigned to its closest oxygen atom within a cutoff distance of 1.2 Å. The OH bond vectors are oriented pointing towards the hydrogen and the water dipole orientation (valid only for water molecules) is represented simply by the vector sum of the two OH bond vectors. The octahedral and tetrahedral assignment is based on the identification of the iron rows and we used the location of the vacancy to distinguish the water on dissociation and the grade of dissociation (see [27]).

To examine the statistical distribution of the orientations, joint probability densities are calculated by making histograms. Specifically, two 2D histograms are generated for the variable pairs ($\cos(\theta)$, $\phi$) and ($\cos(\theta)$, $r$). The probability densities are considered as a function of the cosine of $\theta$ to account for the change in volume element when transforming from Cartesian to spherical coordinates. Finally, we calculated the joint probability distribution of the OH- and H-bond distances, where the H-bond length is defined as the second shortest distance from the hydrogen atom to an oxygen atom (the shortest being the covalent OH-bond involved in the H-bond).

The theoretical framework for calculating the SFG spectra from MD simulations has been developed and applied to calculate the measured resonant frequencies [40–46]. Here, we adopted the approach developed by Khatib and Sulpizi in Ref. [43] in which the imaginary part of the electric response involves the velocity autocorrelation and the derivatives along the OH bond of the polarizability tensor and dipole moment. The details of this calculation are reported in the SI.

## 3 Results and discussion

### Experiment

**$Fe_3O_4$ layer preparation and characterization** Magnetite films of ∼50 nm were prepared by RF-sputtering (see SI for details) and characterized by Raman spectroscopy (Fig. 1a). The Raman spectra at different spots were similar, indicating film homogeneity. Two prominent peaks were observed around ∼685 and ∼1343

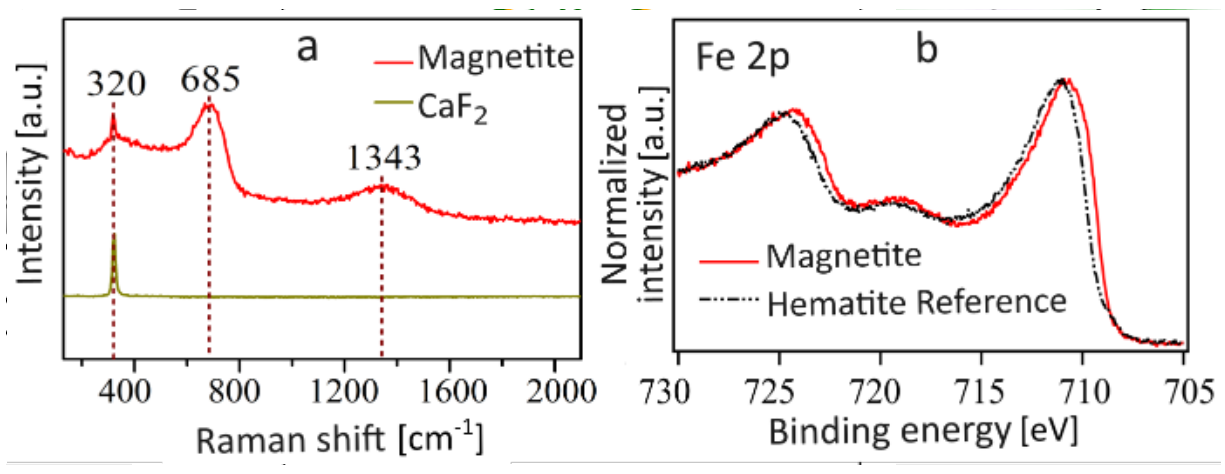

Figure 1: (a) Raman spectra of a magnetite film coated on $CaF_2$ and the bare $CaF_2$ substrate. (b) XPS spectra of the magnetite film in the Fe 2p region.

cm$^{-1}$ and assigned as $A_{1g}$ mode and a two-phonon scattering mode of magnetite, respectively [3, 4]. Another common mode seen for magnetite is around ~330 cm$^{-1}$ representing the $T_{2g}$ mode [3, 4], which overlaps with the $CaF_2$ substrate-specific mode at 320 cm$^{-1}$.

X-ray photoelectron spectroscopy (XPS) measurements of the $Fe_3O_4$ film determined the elemental composition of the near-surface region. Here, the energy scale was calibrated to adventitious carbon at 285.0 eV (see SI). Data from the Fe 2p region are presented in Fig. 1b. Due to the complexity and challenges in fitting the Fe 2p region [47], we compared the $Fe_3O_4$ sample data to a hematite ($\alpha$-$Fe_2O_3$) reference sample (dotted black line) which only contains the $Fe^{3+}$ oxidation state. This allows more easily to discern the contribution of the $Fe^{2+}$ component of $Fe_3O_4$ found in the lower binding energy edge of the 3/2 peak ( ~ 710 eV) [48]. The broadening due to the mixed $Fe^{2+}$ and $Fe^{3+}$ in $Fe_3O_4$ results in the Fe 2p 3/2 (~710 eV) and 1/2 (~724 eV) peaks appearing at slightly lower binding energies compared to hematite [49]. Both samples exhibit the distinct $Fe^{3+}$-related shake-up satellite around 719 eV. The intensity of this peak in the $Fe_3O_4$ data indicates oxidation of the magnetite surface, which is expected as $Fe_2O_3$ is the thermodynamically stable phase under both ultrahigh vacuum [31] and ambient conditions [50]. Details on film fabrication and technicalities of the film characterization can be found in the SI.

**SFG spectroscopy** To unravel the orientation and H-bond strength of water molecules interfacing with magnetite, we performed SFG experiments with the magnetite film in contact with water in the O-H stretch region. Fig. 2a demonstrates the normalized SFG intensity ratio of the magnetite sample in contact with pristine water and pH 11 solution in the OH-stretch region (3000-3750 $cm^{-1}$). The pH 11 measurement serves as a example for a heavily hydroxylated surface. As the magnetite film was observed to generate a position and wavenumber dependent signal, the SFG data were plotted as a ratio with the spectrum measured with $D_2O$ at a fixed position in the sample cell. Since $D_2O$ shows no vibrational features in the measured region, the $Fe_3O_4$-$D_2O$ interface reflects the signal only from the $Fe_3O_4$ film (see reference [33] for details). Hence, an overall SFG ratio value of 1 represents a negligible contribution of the O-H stretching modes. Fig. 2a shows homodyne SFG measurements for pristine $H_2O$ and a pH 11 solution normalized by $D_2O$. As the measured SFG spectrum is a product of Fresnel factors and the $\chi^{(2)}$ response [51, 52] (see SI for details), the true $\chi^{(2)}$ amplitude of the magnetite-$H_2O$ interface is extracted by implementing a Fresnel correction of the $D_2O$ normalized SFG spectra. Fig. 2b illustrates that the Fresnel correction has a major influence on the overall spectral shape and minimizes the broad contribution above 3500 $cm^{-1}$ observed for the uncorrected SFG spectra (Fig. 2a). The corresponding Fresnel factor as function of IR wavenumber can be found in the SI (Fig. S2). This clearly represents the significance of employing the Fresnel correction to SFG spectroscopy data for identifying true spectral features. After Fresnel correction, the SFG ratio for $H_2O$ shows a signal larger than 1 around 3100 $cm^{-1}$ and a signal smaller than 1 above 3300 $cm^{-1}$. The pH 11 spectrum has an intensity above 1 at wavenumbers lower than 3180 $cm^{-1}$ and is below 1 above it. Fig. 2c shows the corresponding measured Im[$\chi^{(2)}$] SFG spectra normalized with $D_2O$ spectra. As explained in the SI (Section 1.8), the signal from the $Fe_3O_4$ film is predominantly real within the experimental phase uncertainty.

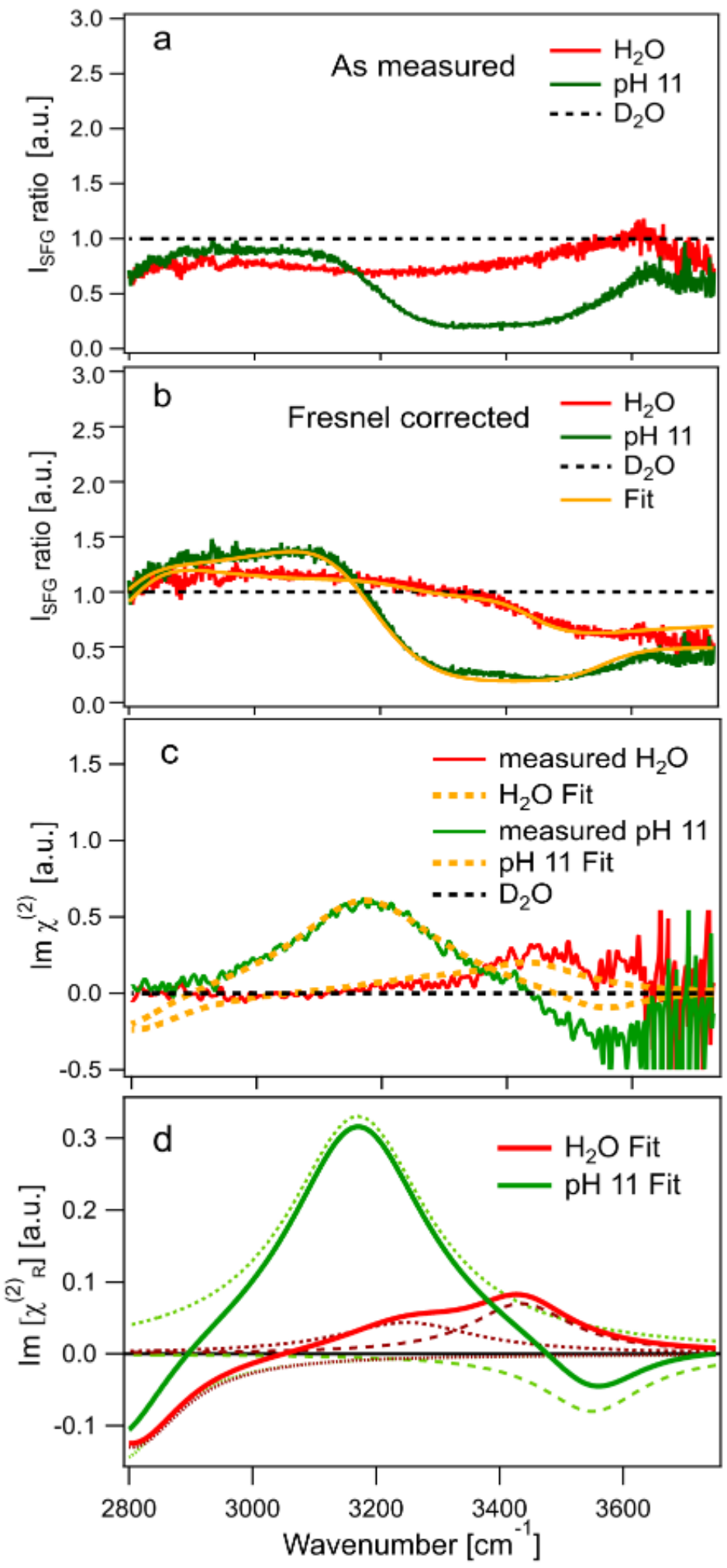

Figure 2: SFG intensity ratio of water-magnetite interface in O-H stretching region for pristine water ($H_2O$, red), and pH 11 (green) as function of IR wavenumber presented (a) as measured and (b) after Fresnel correction. The yellow line represents the fit obtained by the Lorentzian model. (c) Fresnel corrected Im $[\chi^{(2)}]$ from the phase-resolved SFG measurements (solid lines) overlapped with the Im $[\chi^{(2)}]$ amplitude (dotted lines) extracted from the fitted conventional SFG data of $H_2O$ and pH 11, respectively. All spectra are normalized with the $D_2O$ spectrum taken as a reference measurement at the magnetite-$D_2O$ interface. The black dotted line depicts normalized $D_2O$ reference spectrum. (d) Resonant contribution from the fit for $H_2O$ and pH 11, the individual components are shown in dashed lines.

**Insights into the water structure** In the $H_2O$ spectrum (Fig. 2c), the signal is near zero at low wavenumbers, and positive from 3030 $cm^{-1}$ onwards.The positive signal suggests the presence of OH groups pointing with their H-atoms towards the interface. The pH 11 spectrum has a positive signal below roughly 3400 $cm^{-1}$ which indicates O-H groups pointing with their H-atoms to the magnetite surface, while above a small negative signal is observed. To explicitly quantify the intensity variations and relative contribution of the different OH species, we fitted the Fresnel corrected SFG spectra from Fig. 2b by using the model described in the SI and in reference [33]. In short, $\chi^{(2)}$ consists of a nonresonant contribution, taken from the magnetite-$D_2O$ reference spectrum, and a resonant contribution. The resonant susceptibility factor ($\chi_R^{(2)}$) is deduced by using three complex Lorentzian peaks to reconstruct and directly compare experimental data with simulation results. The fit results are depicted as yellow lines in Fig. 2b and c. The Re[$\chi^{(2)}$] can be found in the SI (Fig. S5). A minimum number of Lorentzian components were used to reproduce the experimental spectral shape in Fig. 2. Similar linewidths were used for neutral and alkaline spectra fit to reduce the overall number of fitting parameters. Fig. 2d shows the imaginary part of the resonant contribution obtained from the fit for both the $H_2O$ and pH 11 sample. As the NR contribution is predominantly real within the experimental phase uncertainty (see SI), the difference with panel c is very small only correcting for the small frequency dependence in the NR amplitude. The three individual Lorentzian components are as well added to this panel. For pristine $H_2O$ the result consists of a negative low wavenumber band, and two positive bands at higher wavenumber, representing species with stronger and weaker hydrogen bonding interactions, respectively. The negative (positive) sign indicates that the O-H moieties are pointing with the H atom away from (towards) the interface. For pH 11, the high wavenumber band has an opposite sign and the middle wavenumber band has a significantly increased amplitude compared to pristine water. This is a typical signature for water molecules next to a charged interface [33, 53–56]. As the isoelectric point of magnetite was reported to be between 6.5 and 7[57, 58], the magnetite surface at pH 11 likely acquires a negative charge which explains the substantial signal observed for pH 11 in the lower wavenumber region. Performing salt screening experiments shown in the SI (Fig. S10), indicates that the surface is close to neutral at pH 7, but charged at pH 11.

Similar measurements were performed with both the conventional and phase-resolved SFG in the OD-region at the water-magnetite interface exposed to $D_2O$ and pD 11 solutions. In this case, the measured data is normalized with a measured $H_2O$ reference. The corresponding data interpretation in the OD-region matches well with the OH-region and can be found in the SI (Fig. S3). Moreover, in the SI (Fig. S6) an SFG intensity spectrum for epitaxial grown $Fe_3O_4$ layers with a thickness of 55 nm on a MgO substrate are shown. Details about the layer preparation and characterization can be found in literature [59]. The results are very similar to the layers grown on $CaF_2$.

Table 1: Average populations and standard deviations for each OH species of a 4x4 replicated u.c. surface.

| **Species** | **Mean** | **Sigma** |
|---|---|---|
| dissociated OH | 15.1 | 0.4 |
| 0 dissociation | 1.7 | 0.3 |
| 1 dissociation | 13.4 | 0.2 |
| 2 dissociations | 0.9 | 0.1 |
| remaining OH | 15.0 | 0.4 |
| water on dissociation | 15.2 | 0.3 |
| dipole up | 0.65 | 0.1 |
| dipole middle | 12.8 | 0.2 |
| dipole down | 1.7 | 0.3 |
| OH on dissociation | 0.12 | 0.01 |
| $H_3O$ on dissociation | 0.01 | 0.001 |
| octahedral water | 51.5 | 0.5 |
| tetrahedral water | 49.6 | 0.4 |

## Simulation

The magnetite SCV surface directly exposes two layers to the water molecules. The outermost layer is formed by rows of "octahedral" iron atoms (and oxygen atoms) and the second layer is formed by rows of "tetrahedral" iron atoms, with one iron atom missing per surface unit cell. At this site, a water molecule may dissociate, with one of its hydrogen atoms binding to an oxygen of the magnetite surface. We refer to this species as "dissociated OH". The left hydroxyl group, which we call "remaining OH", adsorbs on an octahedral iron atom [21]. On top of the dissociation site, strongly interacting with the dissociated OH, is a peculiar water molecule, referred to as "water on dissociation". The sketch in Fig. 3 visualizes the above mentioned species.

We analyze the NNP-based simulations of the liquid water/magnetite interface identifying the OH species present in the first water layer (up to 3.8 Å from the surface [27]). The specification of the simulated systems and the details of the analysis can be found in the section Materials and Methods. The average populations of each species are listed in Table 1. On the surfaces of our simulations (consisting of 16 surface u.c.), there are 15.1 dissociated OH and, correspondingly, 15 remaining OH. Of the 16 dissociation sites, with further differentiation, we find that 13.4 host one dissociation, 1.7 are free from dissociation and 0.9 host two dissociations (see Fig. 4). With our identification criteria, we count only 15.2 water on dissociation. The missing 0.8 are mostly water molecules that are too far from their dissociation site (counted as normal tetrahedral water), but can also be in transition states that mediate the hopping process of the dissociated hydrogen (see the values of the OH and $H_3O$ on dissociation) [27, 60]. Interestingly, we observed a correlation between the orientation of the water on dissociation and the number of dissociated OH present at the site (see Table 1 and Fig. 4). More about the specific orientation of the water on dissociation will be discussed in the next paragraph. The number of water molecules adsorbed on the octahedral iron row is 3.2 per unit cell and above tetrahedral rows it is 3.1 per unit cell, in agreement with previous calculations [27].

**The orientation of the OH species at the magnetite interface** To extract complementary information from that available from the SFG experiments, we focus on the orientation of the OH bonds. In Fig. 3a we plot the angular joint probability density ($\cos(\theta)$,$\phi$) of the OH bonds that are not part of water molecules: the dissociated OH in blue and the remaining OH in orange. The blue peak indicates that the dissociated OH bonds consistently point at a polar angle of $\cos(\theta) = 0.71$ ($\theta = 45°$) and perpendicularly to the iron rows, i.e., with an azimuthal angle $\phi = \pi/4$. The intense orange peak shows that OH bonds of the remaining OH, which are adsorbed on octahedral irons, point primarily perpendicularly to the iron rows, slightly away from the surface at roughly $\cos(\theta) = 0.25$ ($\theta = 75.5°$). Deviations of the azimuthal angle $\phi$ from the orthogonal orientation towards the parallel direction are associated with an increase of the polar angle $\theta$. This is likely due to repulsive interactions from nearby hydrogen atoms of water molecules along the rows. The small contribution of the pale orange peak at $\cos(\theta) = 0.25$ ($\theta = 75.5°$) and $\phi = 3\pi/4$, corresponding to an orientation parallel to the rows, arises from intermediate states during hydrogen transfer events between water molecules, implying the delocalization of the hydroxyl groups along the iron rows.

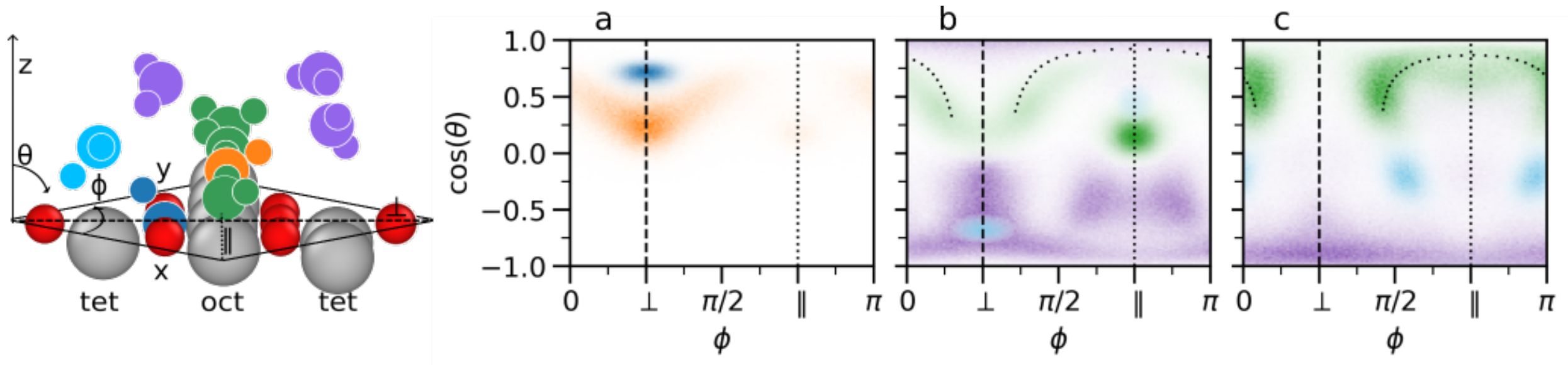

Figure 3: Orientation of the OH species in the first layer at the magnetite interface. The sketch on the left depicts one surface u.c. of the SCV and the various molecular species occurring at the surface. The OH species are in different colors to distinguish between hydrogen coordination and location at the surface: the OH bond dissociated at the surface (blue), the hydroxyl group adsorbed on octahedral iron left after dissociation (orange); the OH of water molecules adsorbed on octahedral iron atoms (green), on top of the dissociation site (light blue) and above tetrahedral iron atoms (purple). The plots show, using the same color scheme, the joint probability densities of the angles $\cos(\theta)$ and $\phi$ of the OH bonds of the hydroxyl groups (a), of the OH bonds of the water molecules (b) and of the dipole moments of the water molecules (c). The angle $\phi$ extends only from 0 to $\pi$ as angles that differ by $\pi$ are identified because of the symmetry of the surface. In each plot, the dashed line and the dotted line indicate the angle $\phi$ corresponding to the directions orthogonal and parallel to the iron rows, as sketched on the left. The dotted curves in plot b and c represent a model for the octahedral OH orientations, as explained in the text.

In Fig. 3b and Fig. 3c, we show the orientation of the OH bonds and the dipole moments of water molecules, respectively. The orientation of the OH bonds of the water on dissociation produces two peaks (light blue in Fig. 3b). One OH bond is orthogonal and points towards the surface with an angle of $\cos(\theta) = -0.67$ ($\theta = 132°$), opposite of the orientation of the dissociated OH (see sketch in Fig. 3); the other OH bond is parallel and points away from the surfaces at $\cos(\theta) = 0.4$ ($\theta = 66.5°$). These orientations form an angle of 105°, which is consistent with the angle of a water molecule. The water dipoles are oriented exactly in the center of the two peaks, at $\cos(\theta) = -0.22$ ($\theta = 102.5°$) shown in light blue in Fig. 3c. Integrating these two OH peaks we obtain a slight discrepancy of populations: 0.51 for the orthogonal peak and 0.44 for the parallel one. The distribution of the dipoles clarifies this mismatch. As expected, the two dipole peaks together contribute to 0.88. The remaining fraction of dipoles is concentrated in a downwards dipole peak (0.09) and an upwards one (0.03) with $\cos(\theta) = \pm 1$ and undefined $\phi$ (unfortunately not visible in Fig. 3c). These two peaks allowed us to reveal the scenarios b) and c) shown in Fig. 4 where the water on dissociation: b) interacts with the surface oxygens with no dissociation (dipole pointing downward), and c) is involved in two H bond with two dissociated OHs (dipole pointing upward). The fractions reflect the probabilities of these configurations to occur, in agreement with previous analysis (see Table 1).

The distribution of the orientation of water molecules on octahedral iron sites (green) reveals that all the OH bonds in this region point away from the surface. Specifically, the distribution shows a distinct peak in the direction parallel to the iron row slightly above the surface at $\cos(\theta) = 0.14$ ($\theta = 82°$), surrounded by a curved cloud. The peak accounts for a fraction of ap-proximately 0.34 of the octahedral OH bonds. Therefore, 0.68 of the water molecules on octahedral irons have one OH bond in the direction of the peak (forming a hydrogen bond network along the iron rows) and the other OH bond somewhere in the pale green cloud surrounding the peak. To test this hypothesis, we compute the curve that a vector that is constrained to form an angle of 104.5° (internal water angle) with the vector oriented as the peak of the distribution would trace if rotated within the top half of the space (i.e., $z > 0$, to avoid penetrating the magnetite slab). The resulting curve and the analogous curve obtained for the dipoles (dotted lines in Fig. 3b and Fig. 3c) are in good agreement with the respective green clouds. The water on the tetrahedral irons (shown in purple in Fig. 3) presents a more disordered orientation with dipoles preferentially oriented with the hydrogens pointing towards the surface.

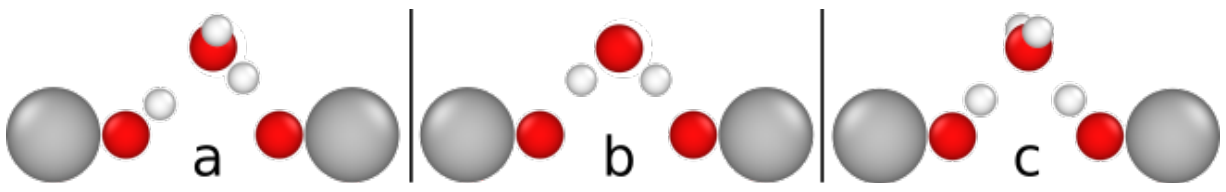

Figure 4: The different orientations of the water on dissociation for the three degrees of dissociation. a) One dissociated OH b) No dissociated OH. c) Two dissociated OHs.

**Bond length analysis** While the orientation of interfacial OH species, quantified by the bond angles $\theta$ and $\phi$, provides valuable insight into the configurations occurring at the surface, it does not quantify the strength of the bonds, that can be characterized by its length. We compute the joint probability density of the OH bond length and the H bond length for the dissociated OH, the remaining OH and the water on the dissociation and compared them with that of bulk water (Fig. 5). As expected, in bulk water (gray in Fig. 5), the covalent OH bond length and the respective H-bond length are inversely correlated [61, 62]. Notably, this inverse relationship holds (and is enhanced) for the interaction of the dissociated OH and the water on the dissociation (blue) and between the latter and the surface oxygen (light blue). The remaining OH (orange) show a much weaker correlation between the OH-bond and the H-bond. To be more quantitative, we calculated the marginal probability densities of the OH bond length and of the H-bond length, shown in the bottom plots in Fig. 5.

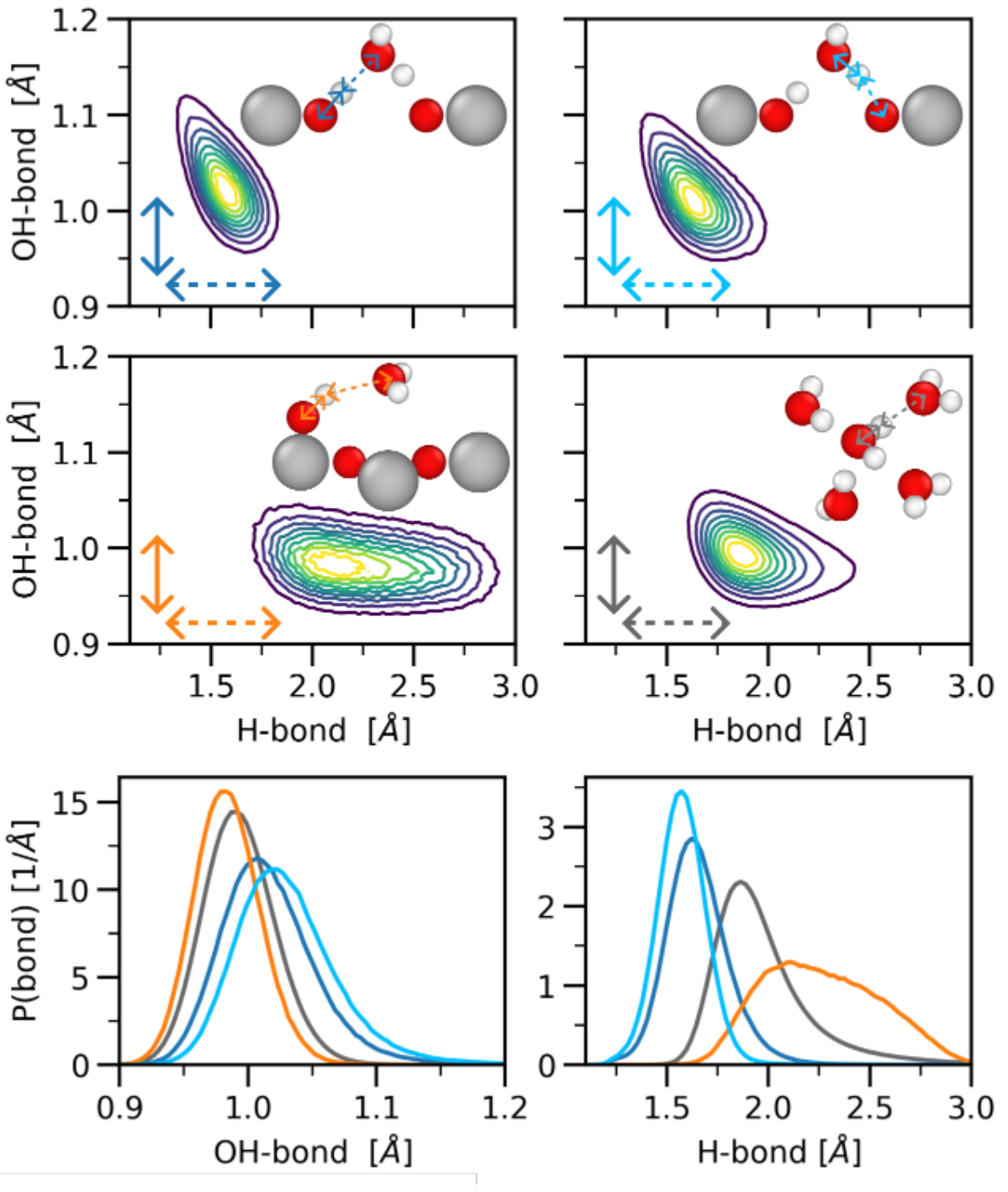

Figure 5: OH species at the magnetite surface from (OH, H)-bond length analysis. The top four panels display the joint probability distribution of the lengths of the OH-bond (solid line) and the H-bond (dashed line) for four different OH species: dissociated at the surface (blue), the OH of water molecules on top of the dissociation site interacting with the surface oxygen (light blue), the hydroxyl group in the first layer adsorbed on octahedral irons (orange) and bulk water (gray). The insets show schematic representations of the corresponding configurations. The level curves are separated by 0.1 of the maximum value. The two plots in the bottom row show the distributions of the OH-bond lengths (left) and the H-bonds lengths (right) for the various species.

Compared to the average OH bond length in bulk water of 0.99 Å, the dissociated OH and the OH of the water molecules on the dissociation site have longer bonds, that average to 1.02 and 1.03 Å, respectively. This difference is due to the very strong H-bonds they form, with average lengths of 1.65 Å and 1.57 Å respectively, compared to the average H-bond length in bulk water 1.97 Å. The hydroxide in the first layer exhibits the shortest OH-bond of 0.98 Å and the longest H-bond of 2.26 Å. These OH bonds, predominantly oriented away from the surface (orange in Fig. 3a), donate a very weak H-bond to the water molecules above the tetrahedral iron, which are located at a larger distance from the surface (below 2.5 Å[27]).

**Stability of the water-magnetite interface** As a prerequisite for the calculatoin of the SFG spectra, we assess the stability of the various species present at the magnetite surface. Each oxygen atom can be unambigously assigned to one of the species discussed above, allowing us to define, for each oxygen atom and each species, a time-dependent binary indicator function [63]:

$$I_i(t) = \begin{cases} 1 & \text{if oxygen } i \text{ is in a given species at time } t \\ 0 & \text{otherwise} \end{cases} \tag{1}$$

Dynamical information of a given OH species is encoded in the time autocorrelation function of its indicator functions. The normalized autocorrelation function at time $t$ can be interpreted as the conditional probability that oxygen $i$ is found in a particular OH species at time $t + t_0$ given that it was already in that species at time $t_0$, i.e. $P(I_i(t_0 + t) = 1|I_i(t_0) = 1)$. The lifetime $\tau$ of a certain OH species is extracted by exponential fitting of the correlation function after averaging over all oxygen atoms, subtracting the long-time plateau and normalizing:

$$C(t) = \frac{\sum_{i=1}^{N} \langle I_i(0)\, I_i(t)\rangle - \langle I_i(0)\rangle\langle I_i(\infty)\rangle}{\sum_{i=1}^{N} \langle I_i(0)\rangle - \langle I_i(0)\rangle\langle I_i(\infty)\rangle} \sim e^{-\frac{t}{\tau}}. \tag{2}$$

The results of this analysis are shown for each OH species in Fig. 6. The lifetime of the dissociated OH is approximately 0.6 ns (Fig. 6a). The dominant physical process captured by this timescale is the hopping of the dissociated proton between two neighboring surface oxygen atoms, mediated by a water molecule adsorbed above the dissociation site. The same hopping process has been experimentally investigated under UHV conditions, where lifetimes on the order of seconds were observed[60]. The remaining OH is also mobile with a comparable lifetime of 0.5 ns. In this case, proton transfer occurs between water molecules adsorbed on octahedral iron sites [27]. In contrast, the water molecule occupying the dissociation site is significantly more stable, exhibiting a lifetime of 1.9 ns. Its dynamics involve transitions between the configurations depicted in Fig. 4b–c, as well as exchange with surrounding water molecules.

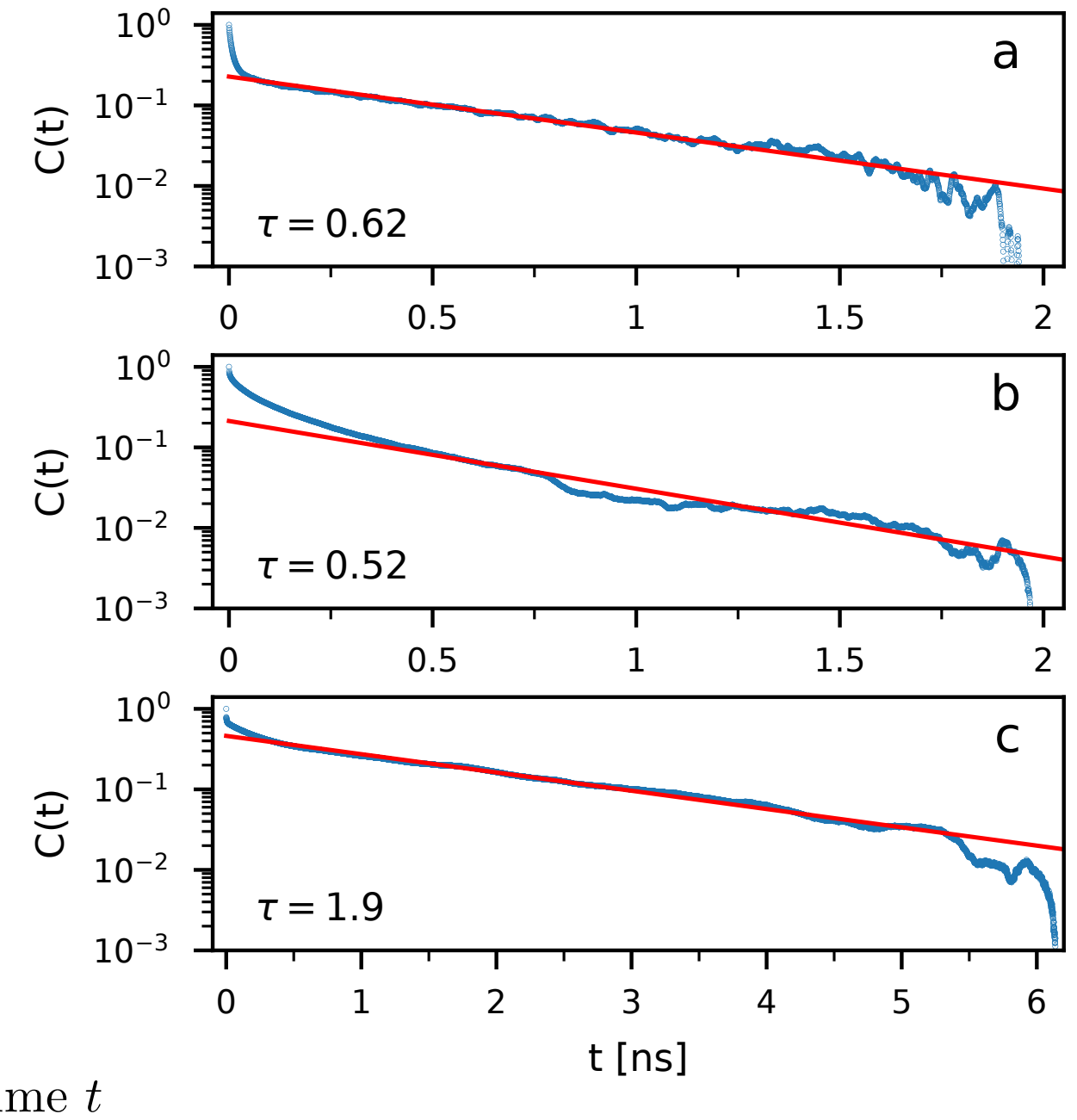

Figure 6: Autocorrelation functions (blue scatter points) and exponential fit (red lines) of the indicator functions for: a) dissociated OH, b) remaining OH and c) water on dissociation interacting with a single dissociation (scenario a) in Fig.4). For all species, the autocorrelation exhibits an abrupt initial decay at very short times. This behavior originates from rapid transient flickering of the indicator functions, which is a consequence of the hard geometric cutoffs used to identify the species. At very long times, subtracting the constant plateau causes the autocorrelation to artificially drop as statistical noise dominates. These short-time and long-time artifacts do not affect the physically meaningful exponential decay and were therefore excluded from the fitting procedure.

The calculated lifetimes highlight the high reactivity of the magnetite surface in contact with water, yet they are long enough to allow the computation of the SFG spectra, which requires lifetimes of at least a few picoseconds.

**The simulated response spectra** To connect the interfacial OH structure to its spectroscopic response, we calculated the SFG spectra resolved by OH species and with successive wa-ter layers included (Fig. 7). The simulated spectra represent the in-plane-averaged resonant response $\mathrm{Im}[\chi^{(2)}_{xxz,R} + \chi^{(2)}_{yyz,R}]/2$, hereafter denoted by $\mathrm{Im}\,\chi^{(2)}$. This quantity corresponds to the experimental SSP geometry under the assumption of an azimuthally isotropic distribution of surface domains. The tensor projection, averaging procedure, and computational details are described in Methods and the Supporting Information.

To relate the simulated spectrum also with our previous bond analysis, a correspondence between the vibrational frequencies $\omega$ of the OH stretch modes and the corresponding bond lengths $r$ can be established. As shown in Fig. 5, our results reveal a statistical anti-correlation between the length of OH-bonds and the H-bonds they are involved with. In bulk water, the OH stretch frequency is known to correlate directly with hydrogen-bond length [61, 62]. Assuming this relationship remains valid at the interface, shorter OH bonds correspond to higher wavenumber features, whereas longer OH bonds give rise to lower frequency contributions.

The total simulated spectrum for the first interfacial water layer (black continuous) is dominated by a negative feature centered near 3200 $\mathrm{cm}^{-1}$, reflecting the prevalence of octahedral water molecules (green), that generate a broad negative spectrum, consistent with the net orientation of their OH groups. This contribution is partially compensated by the narrow positive tetrahedral water feature (purple) that eventually shows up at higher wavenumber, resulting in a net surface-oriented response at around 3500 $\mathrm{cm}^{-1}$. At low wavenumber ($< 3000$ $\mathrm{cm}^{-1}$) the peaks of the spectra of the dissociated OH (blue) and of the downward OH of the water on dissociation (cyan) appear. These species are strongly hydrogen bonded and exhibit opposite average orientations, leading to contributions of opposite sign to $\mathrm{Im}[\chi^{(2)}]$ as expected. The second hydrogen of the water on dissociation produces the negative feature centered near 3300 $\mathrm{cm}^{-1}$. A weaker negative contribution at higher wavenumber (around 3600 $\mathrm{cm}^{-1}$) is likely attributed to the water on dissociation site in configuration b as shown in Fig. 4. A peak at comparable wavenumber is observed for the remaining OH (orange) which is weakly hydrogen bonded and point towards the bulk. The low intensity of this signal reflects the small dipole moment derivative associated with this species (Supporting Information).

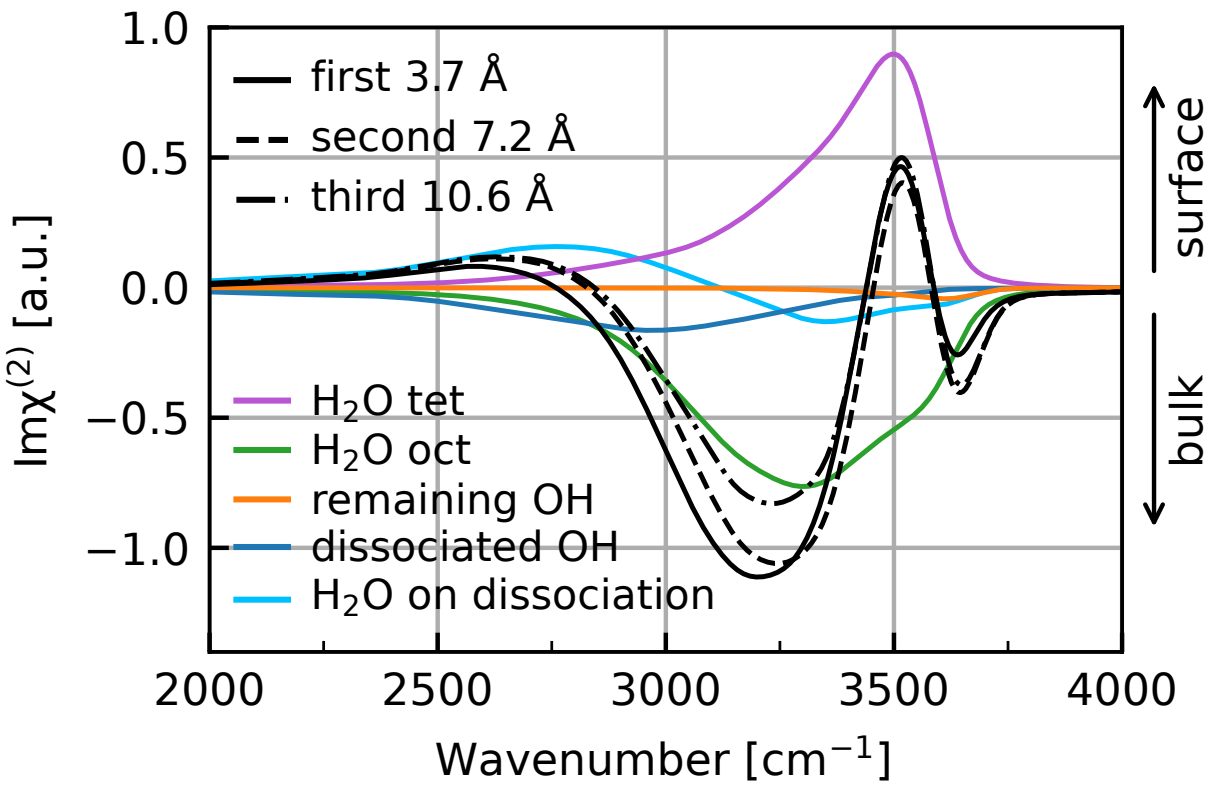

Figure 7: The plot shows the $\mathrm{Im}[\chi^{(2)}]$ of the computed SFG spectra from simulation. The first, second, and third water layer (including water molecules up to 3.7, 7.2, and 10.6 Å from the interface) are plotted in black with different styles. The colors distinguish the contribution of the OH species in the first water layer (the same color scheme as in Fig. 3 is adopted). The axis convention is the same as in the experimental section with positive $\mathrm{Im}[\chi^{(2)}]$ contributions coming from OH bonds pointing towards the magnetite surface.

Including additional water layers in the calculation progressively shifts the spectrum toward positive values of $\mathrm{Im}[\chi^{(2)}]$, demonstrating that water molecules beyond the first layer remain orientationally ordered and contribute coherently to the nonlinear optical response. A detailed analysis of the spatial extent of this orientational ordering and the limitations imposed by the use of a short-range interaction potential is provided in the SI.

## Discussion

A central outcome of our combined experimental and computational analysis is a discrepancy between the experimentally inferred and simulated net orientation of interfacial water molecules at the magnetite–water interface (Fig. 2d and 7). As inferred from the sign and relative intensity of the SFG response, the experiments indicate a dominant population of water molecules with their OH dipoles oriented toward the magnetite surface, whereas the simulations predict the opposite net orientation, with OH groups preferentially pointing with the hydrogen toward the bulk. Assuming the reliability of the simulations—which yield chemically reasonable structures and internally consistent results—this disagreement suggests that the SCV termination, as currently modeled, may not represent the water interfacial structure of the sample measured in the SFG spectroscopy experiments. This possibility is supported by surface X-ray diffraction measurements showing that exposure of $Fe_3O_4(001)$ to water vapour in the mbar pressure regime lifts the surface reconstruction and induces surface roughening [64]. Furthermore, it is well established that there is a rich $Fe_3O_4$ chemistry upon contact with water: $Fe_3O_4$ in contact with liquid water in UHV showed a restructuring of the surface through extensive hydroxylation and formation of an oxyhydoxide phase [23]. In macroscopic amounts of liquid water at ambient pressure selective Fe dissolution might further increase hydroxylation or surface restructuring. Additionally, dissolved oxygen in the liquid phase can lead to oxidation, although relatively slow at room temperature [65].

In particular, the simulated SCV interface may not reproduce the relative populations of OH species that contribute to the experimentally observed Im $[\chi^{(2)}]$. Yet, the simulations reproduce the key spectroscopically active OH species that contribute in the experimental SFG spectra: the low wavenumber negative signature corresponds to the competition of the dissociated OH in strong interaction with the water on top of dissociation; OH stretches whose frequencies and orientational preferences are compatible with the broad positive contribution arise from the water molecules located above the tetrahedral iron rows (panel Fig. 2d). This suggests that the discrepancy does not stem from an incorrect description of individual species, but rather from their relative abundances and orientational weighting in the interfacial ensemble. While the low wavenumber feature is not enough resolved either in the simulated spectrum neither in the experiment, the large opposed signals above 3000 $cm^{-1}$ needs to be accurately discussed.

Within this framework, the SCV simulations suggest that the experimental interface may be characterized by a higher relative contribution of tetrahedral water molecules (with OH groups pointing toward the surface) compared to octahedral water molecules adsorbed on iron sites and oriented toward the bulk. Such a redistribution would directly affect the sign and magnitude of the total Im $[\chi^{(2)}]$ through incomplete cancellation between oppositely oriented OH populations.

One possible structural origin for this imbalance is a partial depletion of octahedral iron adsorption sites under experimental conditions. Aqueous $Fe^{2+}$ release accompanied by the formation of an oxidized, passivating surface layer has been reported for magnetite at pH 1–7 [66]. In the absence of these iron sites, water molecules that would otherwise adsorb on octahedral iron may donate now strong hydrogen bonds to the undercoordinated surface oxygens left after iron removal, thereby contributing positively to the SFG response. Alternatively, the surface in the experiment might be in a so-called oxyhydroxide phase in which a tetrahedral iron flips up and get terminated by two OH groups. Simultaneously two surface oxygens are terminated by the remaining hydrogen atoms. Such a structure has been proposed based on a liquid and ambient-pressure water study on UHV-prepared $Fe_3O_4(001)$ surfaces [23]. In this case water molecules might hydrogen bond with hydrogens pointing to the surface to the oxygen of the oxyhydroxide group enhancing a positive contribution in the SFG spectra. Another, complementary explanation involves the degree of dissociation present at the surface. If the experimental interface contains a lower degree of dissociated OH per dissociation site than assumed in the SCV-based model, the water on dissociation would preferentially orient the dipoles toward the surface (corresponding to scenario b in Fig. 4), enhancing their positive contribution to Im $[\chi^{(2)}]$ in a higher wavenumber window (read also the following pH 11 discussion). Structurally, this scenario can be realized if the magnetite surface would present in the tetrahedral row more dissociation sites (i.e. spots without tetrahedral irons), going back in the direction of a bulk-truncated termination of the magnetite. Theoretical calculations have shown that hydration can alter the relative stability of SCV and bulk-truncated $Fe_3O_4(001)$ surfaces [15]. Although this statement may appear contradictory, it becomes consistent if we consider that the degree of dissociation would greatly depend on the surface electronegativity (i.e. the capability of the surface oxygens to capture hydrogen from water) rather than the bare number of dissociation sites. Consequently, an increase in the number of dissociation sites would lead to a higher ratio of configuration b relative to configuration a in Fig.4.

The argument above provides a natural explanation for the increased SFG intensity observed at pH 11 in Fig. S3b. At this pH condition the surface is expected to be OH-rich and consequently the water would be less prone to dissociate (see also discussion in the SI about the pH 11 measurements in conjunction with speculative simulations of OH-rich surface). The associated spectral feature, centered around 3200 $cm^{-1}$, is also consistent with this assignment, as the OH stretches of water molecules hydrogen-bonded to surface oxygens at dissociation sites are expected to be blue-shifted with respect to the water on dissociation site interacting with one dissociated OH (peak around 2800 $cm^{-1}$). At the same time, under OH-rich conditions such as pH 11, hydroxyl groups adsorbed on octahedral iron sites are expected from simulations to orient toward the bulk, giving rise to higher-frequency OH features. The emergence of such a bulk-oriented OH contribution is indeed observed experimentally as an additional high-frequency feature in panel d of Fig. 2, further supporting this interpretation.

# 4 Conclusions

In this work, we combined surface-specific SFG spectroscopy with reactive NNP-based molecular dynamics simulations to resolve the molecular structure of water and hydroxyl species at the water-magnetite interface. The vibrational response of interfacial water is governed not only by the presence of adsorbed water and hydroxyl groups, but by their relative populations, orientations, and hydrogen-bonding environments. By combining experiment and simulation, we show that these microscopic features can be connected directly to the sign, spectral position, and intensity of the measured nonlinear optical response.

The main experimental finding is that the water-magnetite interface exhibits a pronounced and pH-dependent SFG response in the OH/OD stretching region. At near-neutral pH, salt-screening experiments indicate that the contribution from surface-charge-induced water ordering is negligible, allowing a direct comparison with the computed interfacial response. At pH 11, the enhanced low-wavenumber SFG intensity is consistent with a negatively charged, more hydroxylated interface. These observations demonstrate that the water structure at magnetite is highly sensitive to surface protonation and charge, both of which are key variables in colloidal stability, corrosion, environmental reactivity, and catalytic processes involving iron oxides.

The simulations provide a molecular interpretation of the spectroscopic response. The NNP-MD trajectories reveal a heterogeneous and reactive interfacial hydroxyl network composed of dissociated surface OH groups, hydroxyl groups adsorbed on octahedral iron sites, water molecules bound at dissociation sites, and water molecules above tetrahedral rows. These species differ strongly in orientation, hydrogen-bond strength, and lifetime. The calculated spectra assign the low-wavenumber response to strongly hydrogen-bonded OH species at dissociation sites, while higher-wavenumber contributions arise from more weakly hydrogen-bonded hydroxyls and interfacial water molecules. Compared with earlier UHV, DFT, and classical simulation studies of water on magnetite [12, 13, 15, 27, 67], the present work advances the field by directly connecting surface-sensitive spectra measured under liquid-water conditions to atomistically resolved, reactive interfacial configurations.

A key outcome of the comparison is that the experimentally inferred net orientation of interfacial water differs systematically from that predicted for the simulated SCV-terminated $Fe_3O_4(001)$ surface. Importantly, this discrepancy does not imply that the individual simulated species are unphysical. Rather, the simulations reproduce several spectroscopically active OH motifs observed experimentally, while the total SFG response indicates that their relative populations or orientational weights differ from those of the experimental interface. Thus, the disagreement itself becomes chemically informative: it constrains the possible aqueous termination of magnetite. Our results suggest that the real interface may contain a larger relative contribution from water molecules oriented with their hydrogens toward the surface, possibly due to partial depletion of octahedral iron sites, oxyhydroxide-like motifs, or a lower degree of water dissociation than in the ideal SCV-based model. Such structural motifs are consistent with previous reports that magnetite surfaces can restructure under liquid or near-ambient water conditions [23].

Future work should therefore focus on constructing and testing alternative aqueous magnetite terminations, including partially hydroxylated, iron-deficient, and oxyhydroxide-like surfaces, and on extending reactive simulations to include both explicit long-range electrostatics [68] and other surface termination [15, 23, 64, 66]. Experimentally, systematic measurements as a function of pH, ionic strength, isotope composition, and surface preparation will be essential to separate local interfacial OH signatures from diffuse-layer contributions.

## Author contributions

CD and EHGB designed the project. SR performed the simulations with support of PMdH, YL and CD. HK and MZ did the SFG and Raman spectroscopy experiments and analysis under the supervision of EHGB. HK and MZ produced the RF sputtered magnetite samples and MJC and YD the epitaxial grown ones. ME and GSP were responsible for the XPS measurements and analysis. The manuscript was written with contributions from SR, HK, PMdH, MZ, EHGB and CD. All authors reviewed the article and approved the final version of the manuscript.

## Conflicts of interest

“There are no conflicts to declare”.

## Data availability

The molecular dynamics trajectory of the magnetite water interface is available on the online repository Zenodo, DOI: 10.5281/zenodo.15658131.

## Acknowledgement

We thank Markus Sauer from the Analytical Instrumentation Center, TU Wien, for conducting the XPS measurements. Moritz Eder acknowledges funding from the EU Marie Sklodowska-Curie Actions of Horizon-MSCA-2022-PF-01 (Project 101103731 - SCI-PHI). M.J.C. and Y.D. were supported by the U.S. Department of Energy (DOE) Office of Science, Basic Energy Sciences, Materials Sciences and Engineering Division, Synthesis and Processing Science Program, FWP 10122. The authors also acknowledge the use of facilities at the Faculty Center for Nano Structure Research, University of Vienna, Austria. The computational results presented here have been obtained using the Vienna Scientific Cluster (VSC). This research was fully funded by the Austrian Science Fund (FWF) 10.55776/F8100. 

# Supporting Information

Towards decoding the structure of water-magnetite interface

Salvatore Romano[1,2,†], Harsharan Kaur[3,†], Pablo Montero de Hijes[1], Moritz Zelenka[3,6], Moritz Eder[4], Min-Ju Choi[8], Yingge Du[8], Gareth S. Parkinson[4], Yair Litman[7], Ellen H. G. Backus[3,6,*], Christoph Dellago[1,5,*]

[1]University of Vienna, Faculty of Physics, Boltzmanngasse 5, 1090 Vienna, Austria
[2]University of Vienna, Vienna Doctoral School in Physics (VDSP), Boltzmanngasse 5, 1090 Vienna, Austria
[3]University of Vienna, Faculty of Chemistry, Institute of Physical Chemistry, Währinger Str. 42, 1090 Vienna, Austria
[4]Institute of Applied Physics, TU Wien, Wiedner Hauptstrasse 8-10/134, 1040 Vienna, Austria
[5]Research Platform on Accelerating Photoreaction Discovery (ViRAPID), University of Vienna, 1090 Vienna, Austria
[6]University of Vienna, Vienna Doctoral School in Chemistry (DoSChem), Währinger Str. 42, 1090 Vienna, Austria
[7]Max Planck Institute for Polymer Research, Mainz, Germany
[8]Physical Sciences Division, Pacific Northwest National Laboratory, Richland, WA 99352, USA

† These authors contributed equally.
* Corresponding authors.

# 1 Experiment

## 1.1 Preparation of magnetite thin-films

The magnetite ($Fe_3O_4$) films were prepared by radio-frequency (RF) magnetron sputtering (Korvus HEX, Korvus Technologies) of a commercially procured magnetite sputter target having a diameter of 50.8 mm and a thickness of 2 mm (Evochem Advanced Materials GmbH, Germany, non-bonded, 99.9% purity). A calcium fluoride ($CaF_2$) window (25 mm diameter and 2 mm thickness, Crystal GmbH), used as a substrate for preparing magnetite films, was annealed at 500 °C for 2 hours before deposition. The annealed $CaF_2$ substrate was mounted with Kapton tape on a sample holder within the chamber assembly. The deposition rate was constantly monitored by the quartz crystal microbalance (QCM) installed within the chamber. The deposition of magnetite was performed at 50 W of RF power and a chamber vacuum pressure of 4 x $10^{-4}$ mbar which is regulated by a multi-channel flow controller for Argon gas. Spectroscopic ellipsometry (Sentech, SENpro) with an incident angle of 70° revealed a film thickness of $52.7 \pm 3$ nm using a Cauchy layer model.

## 1.2 Film characterizations

The homogeneity of magnetite samples was inspected with a confocal Raman microscope (WITech alpha 300 ARS). Optical microscopy images were performed with both 10x (0.25 numerical aperture (NA)) and 100x (0.9 NA) objective lenses (EC Epiplan-Neofluar DIC lens, Carl Zeiss AG). The spectral acquisition was done with a 532 nm diode laser as the excitation source at 5 mW power which was below the determined damage threshold. The Raman spectroscopy measurements were done with the 100x objective lens. A 600 g/mm grating (BLZ = 500 nm) was used to disperse the light, which was then detected using a CCD camera. The spectra were collected in the range of 100-4000 $cm^{-1}$ with 10 spectral accumulations integrated over 10 seconds. The data extraction and analysis were performed using the WITec Project 5.3 software. Raman microscopy was also used to routinely check the intactness of the $Fe_3O_4$ layer after SFG spectroscopy, an in detail discussion of the layer stability is given in Section 1.12.

XPS measurements were carried out on a PHI Versa Probe III-spectrometer equipped with a monochromatic Al-K$\alpha$ X-ray source and a hemispherical analyser (acceptance angle: ±20°). Pass energies of 140 eV and 27 eV as well as step widths of 0.5 eV and 0.05 eV were used for survey and detail spectra, respectively (Excitation energy: 1486.6 eV Beam energy and spot size: 50 W onto 200 $\mu$m; Mean electron take-off angle: 45° to sample surface normal; Base pressure: $< 8x10^{-10}$ mbar, Pressure during measurements: $< 1x10^{-8}$ mbar). Samples were mounted on double-sided polymer tape. Electronic and ionic charge compensation was used for all measurements (automatized as provided by PHI). The binding energy (BE) scale and intensity were calibrated by using methods described in ISO15472, ISO21270, and ISO24237. The analysis depth is typically around 7-10 nm. Data analysis was performed using CASA XPS and Multipak software packages, employing transmission corrections, Shirley or Tougaard backgrounds (Shirley, 1972; Tougaard, 1997) and customized Wagner sensitivity factors (Wagner, 1983).

## 1.3 XPS analysis

In the O 1s region (Fig. 1a), a lattice O 1s peak appears at 529.8 eV with a shoulder at 531.2 eV, attributed to hydroxyl groups (Kraushofer et al., 2019) and oxygen from adsorbed contaminants. This contamination is also visible in the C 1s data (Fig. 1b) as a peak at 288.5 eV, consistent with adsorbed formate (Gamba et al., 2015). Formic and acetic acids are known to have a strong tendency for dissociative adsorption on metal oxides (Balajka et al., 2018).

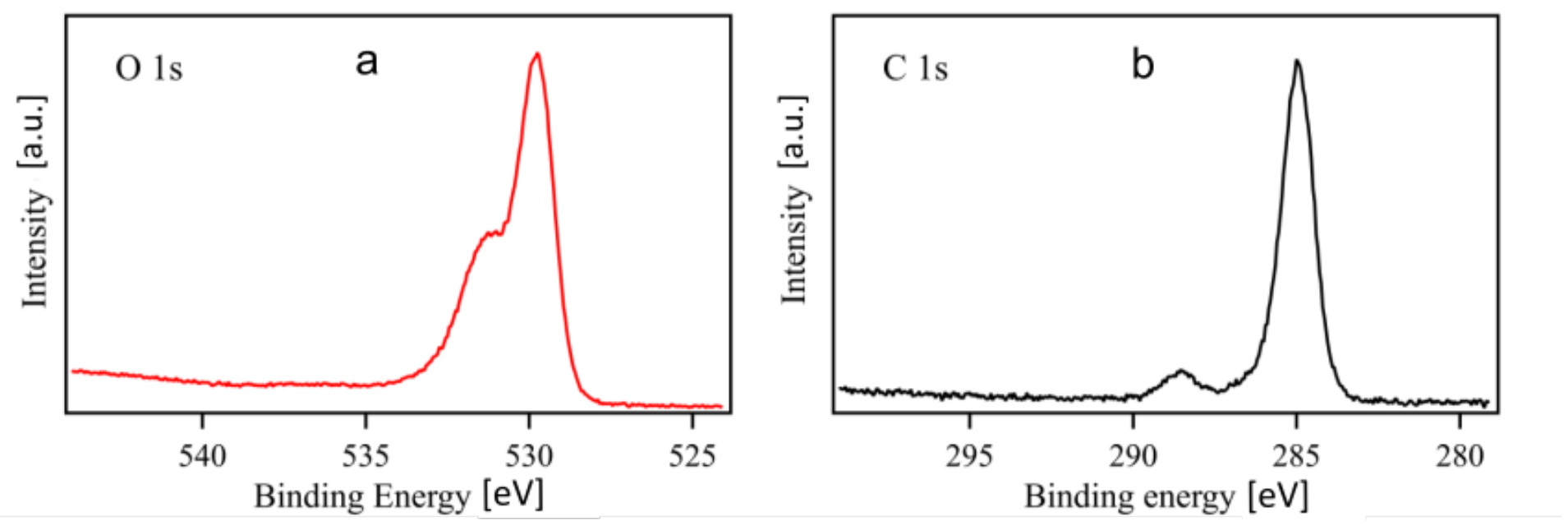

Figure 1: XPS data in (a) O 1s region and (b) C 1s region.

## 1.4 Solution preparation and measurement geometry

All used solutions were equilibrated for several hours in ambient air prior to measurements and had a temperature of 20°C. Ultrapure water (Merck Mili-Q, 18.2 MΩ.cm at 25 °C) and deuterated water ($D_2O$, Deutero GmbH, Kastellaun, Germany, 99.9%) were used for the sample preparation. Solutions with their pH value adjusted to pH 7 or 11 were prepared using a 1 M NaOH (>99.5%, Chemsolute) for both of $H_2O$ and $D_2O$. Solutions with adjusted ionic strength were prepared by addition of NaCl ($\geq$ 99%, ACS reagent, Sigma-Aldrich). The solution pH was measured using a pH meter (FiveEasyPlus, Mettler Toledo). Test measurements of solutions before and after contact with $Fe_3O_4$ layers in our measurement cell showed that no statistically significant pH drift is expected during the SFG spectroscopy measurements.

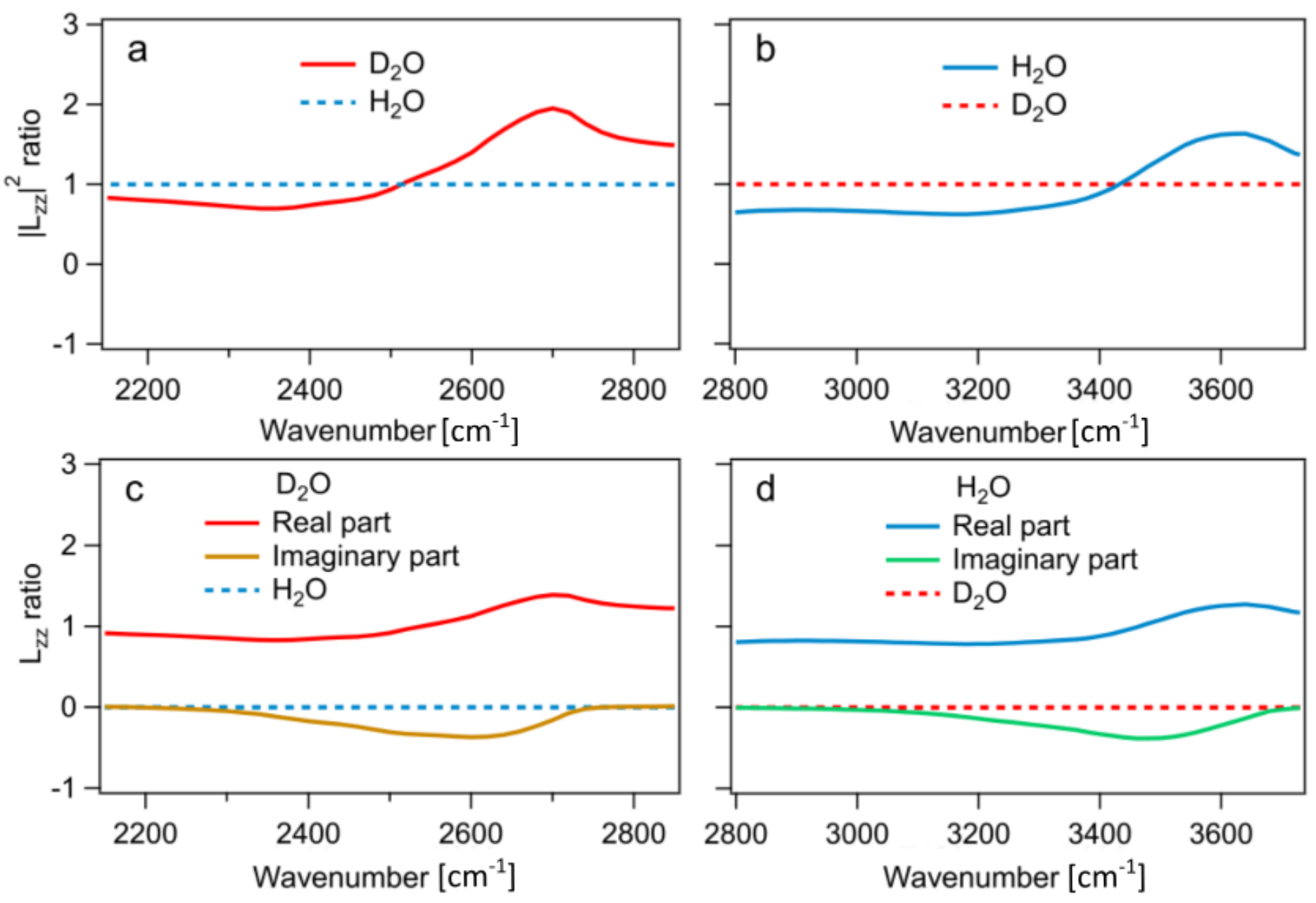

Figure 2: Fresnel factor calculations for the (a,b) intensity ($|L_{zz}|^2$) ratio, and (c,d) the ratio of the real and imaginary part of $L_{zz}$ plotted as a function of IR wavenumber in OD (2150-2850 cm$^{-1}$) and OH (2800-3750 cm$^{-1}$) stretching region, respectively, by considering $D_2O$ and $H_2O$ as a subphase. The data in OD and OH region are normalized with $H_2O$ and $D_2O$ as a reference, respectively, marked as a dotted line in each plot.

In order to probe the behavior of water interfacing the magnetite film, a steel-based sample cell with a flow cell geometry is employed (Zelenka and Backus, 2025). The flow-cell allows to probe the same spot of the magnetite film after solution exchange. The SFG data were collected in static conditions. The flow cell is rinsed with pristine water in between the solution exchange. The pump utilized for this geometry is an analog L/S Masterflex attached to Tygon 2011 tubes (inner diameter = 4.8 mm, Cole-Parmer).

## 1.5 VSFG data analysis

The acquired SFG intensity was subtracted by a background measured by blocking the IR beam and the frequency is calibrated by measuring an SFG signal of gold with a thin polystyrene film placed in the IR path. The SFG data was analyzed using the Igor Pro 8.04 software. The heterodyne measurements of both the sample and reference in the frequency domain were inverse Fourier-transformed into the time domain to choose one of the cross terms, and then Fourier-transformed back into the frequency domain. The normalized phase-resolved spectra were obtained by dividing the spectra of (deuterated) water pH solutions with the reference $D_2O$ ($H_2O$) spectrum. The sign is chosen to have at low frequency a positive signal for high pH, indicating hydrogens pointing towards the interface, based on the assumption that the surface is charged negatively at high pH.

The effective SFG intensity ($I^{SFG}_{eff,SSP}$) in SSP polarization mode is measured as the product of the modulus square of the complex second-order non-linear susceptibility $\chi^{(2)}_{yyz}$ term with the incident intensities of the visible ($I_{VIS}$) and IR ($I_{IR}$) beams, represented as: (Backus et al., 2012; Wang et al., 2019)

$$I^{SFG}_{eff,SSP} \propto |\sum_q (L^n_{yy}(\omega_{SFG}) L^n_{yy}(\omega_{VIS}) L^n_{zz}(\omega_{IR})) \sin\theta_{IR} \chi^{(2)}_{yyz,SSP}|^2 \sec^2\theta_{SFG} I_{VIS} I_{IR} \tag{1}$$

Here, the $L_{yy}$ and $L_{zz}$ are the Fresnel factor components which denote the magnitude of the local surface electric field. The $L^n(\omega)$ component is governed by the wavenumber-dependent complex refractive index (n) of the different interfacial media. The second-order non-linear susceptibility $\chi^{(2)}_{yyz,SSP}$ in the above equation represents the macroscopic average of the resonant response of the surface-bound coordinate system and $\theta_{IR}$ and $\theta_{SFG}$ are the incident and the reflected angles of the IR and SFG beam, respectively. As the SFG spectrum is very different for pristine water (or $D_2O$) and pH 11 (or pD 11), the measured SFG signal originates from the aqueous side of the oxide-water (or $D_2O$) interface. As such, the $L_{zz}$ component is: (Backus et al., 2012; Wang et al., 2019)

$$L_{zz}(\omega) = \frac{t^P_{12}}{1 + r^P_{12} r^P_{23} e^{2i\beta}} (1 + r^P_{23}) \frac{n_1 n_2}{n^2_{interface}} \tag{2}$$

In the above equation, $n_1$ and $n_2$ denote the refractive index of $CaF_2$ (medium 1) and $Fe_3O_4$ (medium 2), respectively. Here, $\beta$ depicts a phase difference factor, $r^P_{ij}$ and $t^P_{ij}$ are the linear reflection and transmission coefficients at the interface between medium $i$ and $j$ for the p-polarized IR beam, respectively ((Backus et al., 2012)). The term $n_{interface}$ is considered equivalent to the refractive index of water or $D_2O$ (Backus et al., 2012). The relative intensity and magnitude ratio of the calculated Fresnel factor $L_{zz}$ with $D_2O$ or $H_2O$ as subphase is plotted as a function of IR wavenumber in both OD- and OH-stretching regions, respectively, as demonstrated in Fig. 2.

For calculating the Fresnel Factors a film thickness of 50 nm and an incident angle at the air-$CaF_2$ interface of 40 degrees with respect to the surface normal was used. The refractive index for $H_2O$ and $D_2O$ are taken from (Bertie et al., 1989). The refractive index from the solids $CaF_2$ and $Fe_3O_4$ are taken from (Li, 1980) and (Querry, 1985), respectively.

To explicitly quantify the intensity variations and relative contribution of the different OH species, we fitted the Fresnel corrected SFG spectra by using the model described in reference (Buessler et al., 2023). However, in the present work, the overall signal for the magnetite-$D_2O$ reference (representing the $\chi^{(2)}_{NR}$ signal) was found to decrease in every measurement which may be attributed to the dissolution of the film with every solution exchange. The phase is stable (see Fig. S5). Therefore, we modified the fit equation with a scaling factor $F$. Considering the signal normalization with $D_2O$, the fitting equation used is expressed as:

$$\frac{I_{H_2O}}{I_{D_2O}} \propto \frac{|\chi^{(2)}_{yyz,H_2O}|^2}{|\chi^{(2)}_{yyz,D_2O}|^2} = \frac{|F * \chi^{(2)}_{NR} + \chi^{(2)}_{R}|^2}{|\chi^{(2)}_{NR}|^2} \tag{3}$$

The complex second-order non-linear susceptibility $\chi^{(2)}$ contains a resonant $\chi^{(2)}_R$ and a non-resonant $\chi^{(2)}_{NR}$ term. The resonant part is fitted as the sum of Lorentz peaks and is given by: (Backus et al., 2012; Buessler et al., 2023; Wang et al., 2019)

$$\begin{aligned} \chi^{(2)} &= \chi^{(2)}_{NR} + \chi^{(2)}_{R} \\ &= A_{NR}(\omega) e^{i\phi_{NR}} + \sum_q \frac{A_q}{\omega_{IR} - \omega_q - i\Gamma_q} \end{aligned} \tag{4}$$

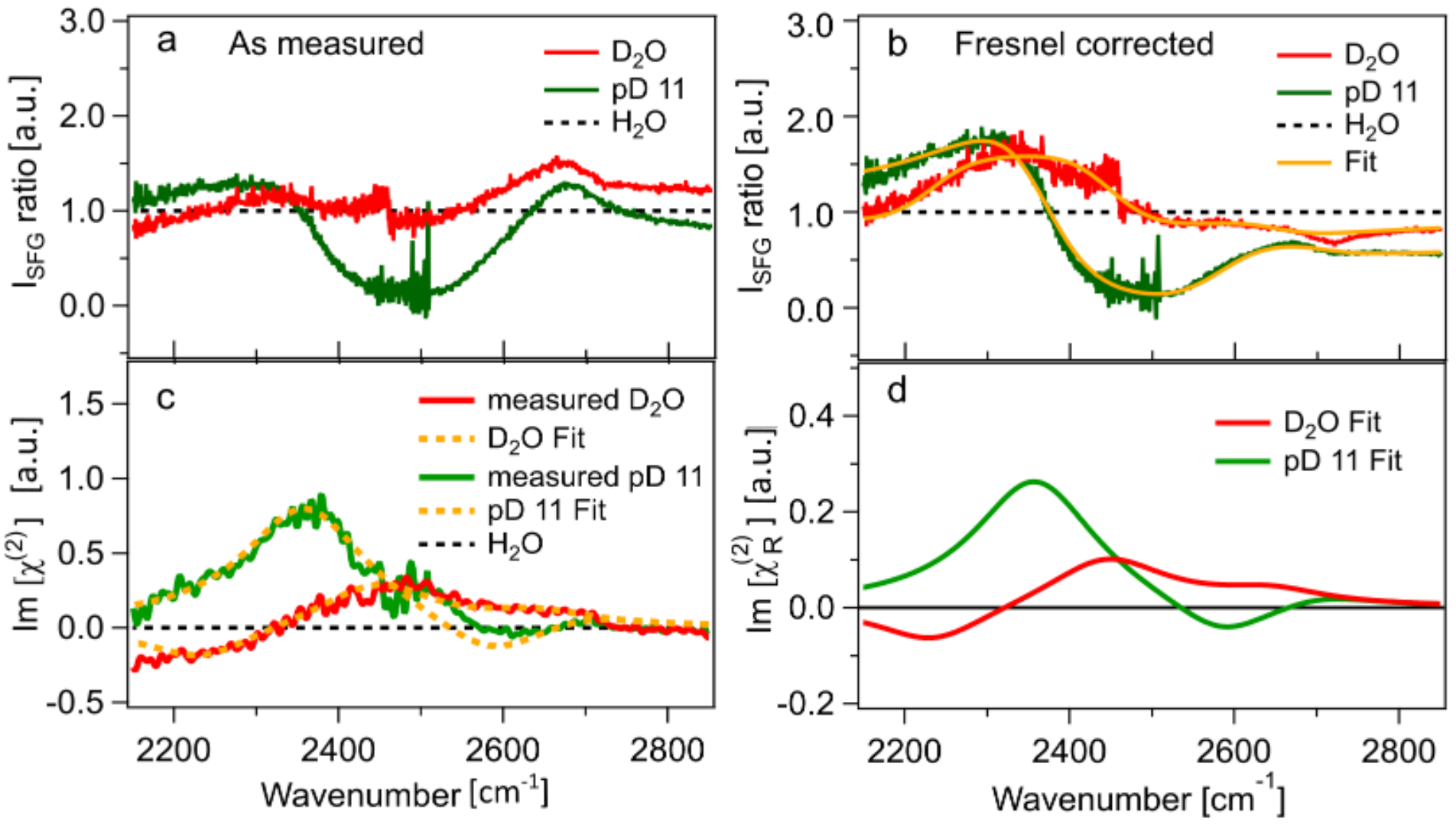

Figure 3: SFG intensity ($I_{SFG}$) ratio of the magnetite-$D_2O$ interface in the O-D stretching region for deuterated water ($D_2O$, red), and pD 11 (green) as function of the IR wavenumber (a) as measured and (b) after Fresnel correction. The yellow line represents the fit obtained by the Lorentzian model. (c) Fresnel corrected Im $[\chi^{(2)}]$ from the phase-resolved SFG measurements (solid lines) overlapped with the Im $[\chi^{(2)}]$ amplitude (dotted lines) extracted from the fitted conventional SFG data of $D_2O$ and pD 11, respectively. All spectra are normalized with the spectrum from the water ($H_2O$)-magnetite interface. The black dotted line depicts the normalized $H_2O$ reference spectrum. (d) Resonant contribution from the fit for $D_2O$ and pD 11.

Here, the terms $A_q$, $\omega_q$, and $\Gamma_q$ designate the amplitude, resonant wavenumber, and full-width half maxima (FWHM) of the $q$-th vibrational mode of the molecular species, respectively. The non-resonant part of the susceptibility is represented by its non-resonant amplitude $A_{NR}$ and the non-resonant phase term $\phi_{NR}$. For current analysis, $A_{NR}$ is determined from the SFG intensity of the magnetite-$D_2O$ spectrum (for OH-stretching region) or magnetite-$H_2O$ spectrum (for OD-stretching region), while $\phi_{NR}$ (used as global fit parameter) is zero. The fit parameters are summarized in Table S1.

## 1.6 Experimental results: OD region

Fig. 3a demonstrates the normalized SFG intensity ratio of the magnetite sample in contact with pristine $D_2O$ and pD 11 solution in the OD-stretch region. The regions 2150-2450 cm$^{-1}$ and 2450-2850 cm$^{-1}$ were measured independently and then combined as a single plot representing a broad measurement range of 2150-2850 cm$^{-1}$. Referring to the similar analysis scheme as mentioned in the main text, the resonant susceptibility factor is deduced by considering three complex Lorentzians. For pristine $D_2O$, these are centered at 2235, 2445, and 2645 cm$^{-1}$. The yellow lines in Fig. 3b and c show the fit results. Similarly, the peak contributions for pD 11 are centered at 2358, 2585, and 2703 cm$^{-1}$ (see Table S1). Overall, the pD 11 spectrum in Fig. 3a and 3b demonstrates more prominent features than the $D_2O$ spectrum. The spectral fit depicted in Fig. 3c and Fig. 3d (only resonant contribution) demonstrates a stronger signal for pD 11 than $D_2O$ in the low- and mid-wavenumber region. This is a typical signature for water molecules next to the charged interface (Buessler et al., 2023; Rehl et al., 2019; Darlington et al., 2017; Ong et al., 1992; Rehl et al., 2022) as also mentioned in the main text. For pD 11, Im $[\chi^{(2)}]$ in the low- and mid-wavenumber region has a positive sign, while the high-wavenumber peak has a negative sign, while the opposite is true for the $D_2O$ spectrum. Here, the positive sign depicts the orientation of water molecules with their deuterium atoms facing towards the magnetite surface, while the negative sign reflects the water species with their deuterium atoms pointing towards the bulk (away from the surface). The assignments observed in the OD region match well with the OH region results (discussed in the main text). However, it is relevant to mention here that the probing of IR wavenumber (into the higher frequency regime) was limited in the OH region owing to the overlapping contribution from the moisture in the surroundings during the experiments, which resulted in a noisy spectrum. The parameters used to model the vibrational features of the $Fe_3O_4$-$D_2O$ interfaces are depicted as well in Table S1. Furthermore, we plotted the resonant SFG response in Figure S4 obtained from fitting as described above to further illustrate the comparability of the information gained from the two interfaces. In addition to the total fitted resonant SFG contribution, a deconvolution showing the single resonances contributing to the spectra is depicted.

Table 1: Fitting parameters used for SFG measurements of $D_2O$, pD 11, $H_2O$, and pH 11, at the magnetite-$D_2O$ (or $H_2O$) interface. Here, A is the amplitude, $\omega$ describes the center wavenumber, $\Gamma$ represents FWHM, and factor 'F' represents the scaling factor.

| | $D_2O$ | pD 11 | $H_2O$ | pH 11 |
|---|---|---|---|---|
| Factor (F) | 1.0 | 0.9 | 0.9 | 0.8 |
| $A_1$ | -7 | – | -13 | -15 |
| $\omega_1$ | 2235 | – | 2804 | 2780 |
| $\Gamma_1$ | 173 | – | 200 | 200 |
| $A_2$ | 10 | 25 | 6 | 45 |
| $\omega_2$ | 2445 | 2358 | 3250 | 3171 |
| $\Gamma_2$ | 184 | 184 | 275 | 275 |
| $A_3$ | 2 | -7 | 7 | -8 |
| $\omega_3$ | 2645 | 2585 | 3435 | 3550 |
| $\Gamma_3$ | 163 | 163 | 200 | 200 |
| $A_4$ | – | 2 | – | – |
| $\omega_4$ | – | 2703 | – | – |
| $\Gamma_4$ | – | 163 | – | – |

Figure 5 shows the Re $[\chi^{(2)}]$ as well as the intensity spectrum reconstructed from the phase resolved spectra.

## 1.7 Experimental results: Phase stability during measurements

The nonresonant signal from the $Fe_3O_4$ layer showed mostly a decreasing trend during measurement series. Figure S6 shows four exemplary measurements of the $Fe_3O_4$-$D_2O$ interface at different time points. The data is normalized to the $Fe_3O_4$-$D_2O$ spectrum at time 0 h to highlight the relative changes of the signal. Figure S6(a) depicts the magnitude of the complex signals. Clearly, there is a decreasing trend observable, which shows that the nonresonant response of the $Fe_3O_4$-$D_2O$ interface gets weaker during the measurement series. Contrary, the respective phase of the complex phase-resolved signal stayed within the experimental uncertainty constant, changing less than 0.2 rad which is around the uncertainty for the phase we obtained by comparing the phase shift of the $Fe_3O_4$-$D_2O$ for different measurement days. Therefore, we fitted our SFG spectroscopy data with a constant phase and introduced a scaling factor F to account for the decreasing trend of the nonresonant signal during a measurement series (see section 1.5).

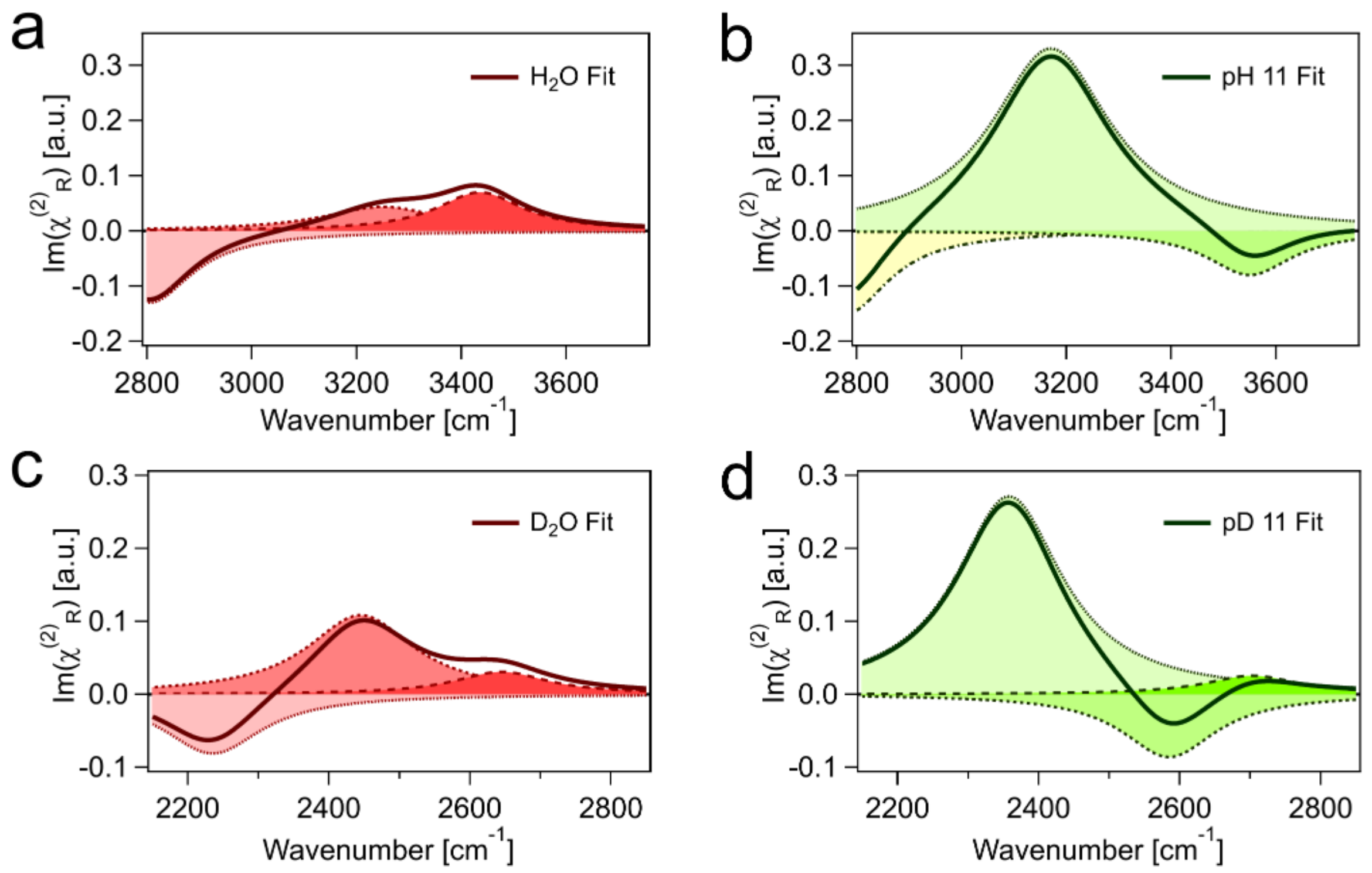

Figure 4: Imaginary part of the fitted resonant SFG response with the parameters shown in Table 1. The single peaks contributing to the total response are shown next to the total contribution. The $Fe_3O_4$-$H_2O$ is shown in (a) at neutral pH and (b) at pH 11, while the $Fe_3O_4$-$D_2O$ interface is depicted in (c) at neutral pH and (d) at pD 11.

## 1.8 Experimental results: Determining the phase of the nonresonant $Fe_3O_4$-$D_2O$ SFG signal

The resonant and nonresonant sum frequency generation (SFG) contributions have a specific phase with respect to the incident laser radiation. We considered different phases of the resonant and nonresonant contributions in the data analysis by introducing a relative phase in the fitting Equation S4. Usually, the nonresonant contribution is considered (mostly) real-valued. However, for an absorptive, magnetic material like $Fe_3O_4$ this does not need to hold true and a complex nonresonant contribution can alter the obtained fit values and consequently the conclusions about the underlying molecular structure. Using the established phase relationship of the nonresonant signal from z-cut quartz which is considered purely imaginary (Hu et al., 2019), we were able to determine the relative phase of the strong nonresonant signal of $Fe_3O_4$ and the resonant signal of the interfacial water molecules similarly to Ref. (Hosseinpour et al., 2017) as follows:

Phase-resolved SFG spectroscopy measurements of a z-cut quartz cube, a 2 mm thick $CaF_2$ substrate coated with a $>$ 100 nm thick gold layer and a 2 mm thick $CaF_2$ substrate coated with an around 50 nm thick $Fe_3O_4$ layer were compared. For the comparison of the quartz and gold signals, the incident beams were traveling through air before reaching the quartz or gold surface. In the consecutive comparison of gold to $Fe_3O_4$, the infrared and visible beams passed first through the $CaF_2$ substrates before reaching the interface with the respective layer, equivalent to the experiments from the main text. Furthermore, the $Fe_3O_4$ layer was brought into contact with $D_2O$ to have comparable refractive indices to the water structure measurements from the main text, without having a resonant signal in the probed frequency region from 3000-3600 $cm^{-1}$. All spectra were corrected by the respective Fresnel factors as described in Refs.

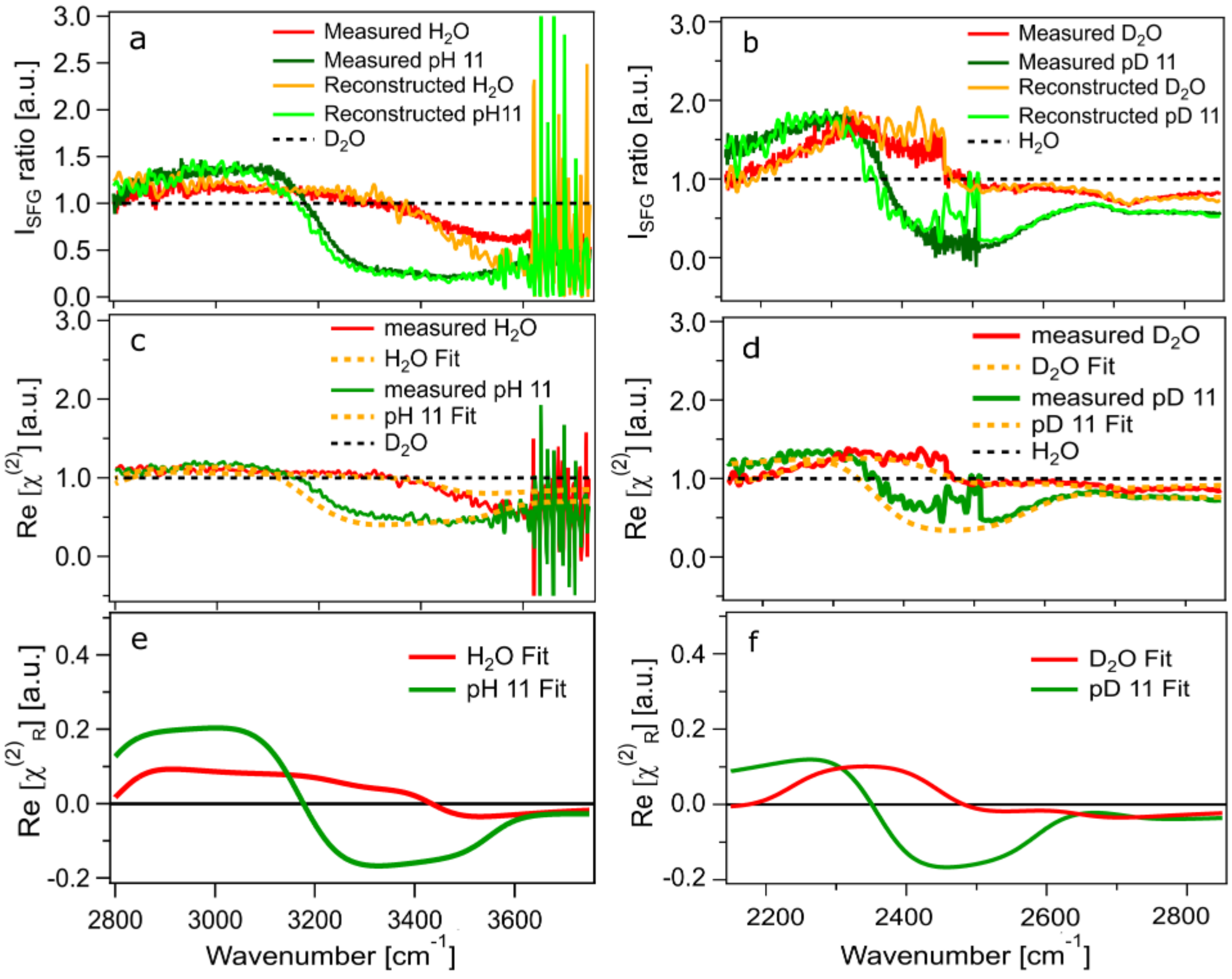

Figure 5: Comparison of conventional SFG spectrum and reconstructed SFG intensity from phase-resolved SFG spectroscopy for (a) the $Fe_3O_4$-$H_2O$ and (b) the $Fe_3O_4$-$D_2O$ interface. (c), (d), (e), (f) ...

(Zhuang et al., 1999; Backus et al., 2012). To evaluate the consistency of our measurements, we probed different $CaF_2$ substrates coated with gold with slightly varying thicknesses and the reproducibility of the quartz measurement. For both, the phase was within our experimental uncertainty unaltered during the tests. For the used $Fe_3O_4$ layer on $CaF_2$ we obtained a standard deviation of 0.2 rad for the average phase from 3 different days.

As pointed out in Ref. (Hosseinpour et al., 2017), both the Fresnel factors $F$ and the reflectivity $r$ of the local oscillator at the different interfaces needs to be incorporated in the spectral analysis, since they lead to additional phase shifts:

$$\chi^{(2)}_{norm.} = \frac{F_{ref}}{F_{sample}} \cdot \frac{r^*_{ref}}{r^*_{sample}} \cdot \chi^{(2)}_{uncorr.norm.} \qquad (5)$$

$\chi^{(2)}_{uncorr.norm.}$ corresponds to the $\chi^{(2)}$ obtained after normalizing the spectra with a reference before correcting the data with the respective Fresnel Factors and reflectivities to get $\chi^{(2)}_{norm.}$. To obtain the relative phase of the gold layer with respect to quartz, we processed the spectra as described in Ref. (Nihonyanagi et al., 2009) using quartz in the x-direction, as defined in Ref. (Hu et al., 2019), for the normalization. The resulting relative phases are depicted in Figure S7A. All spectra are within the measurement uncertainty spectrally flat in the probed frequency range. The gold layer had a phase shift of around -1 rad relative to the quartz. For comparison and to showcase the importance of correcting for the reflectivity, we also plotted the phase shift of the gold signal only corrected with the Fresnel factors, which is closer to a phase shift of $\pi/2$ rad.

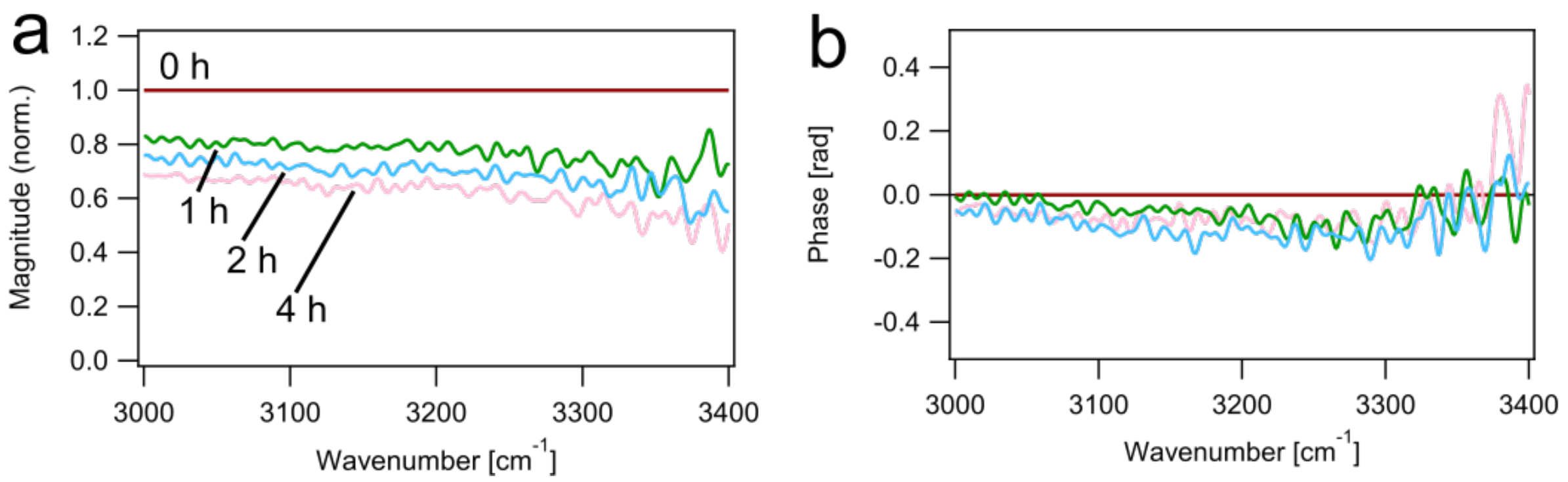

Figure 6: Selected phase-resolved SFG measurements of the $Fe_3O_4$-$D_2O$ interface in the OH-stretch region from a 4 h measurement series. The data is normalized with the first measurement at 0 h for better comparability of the trend. The graphs depict (a) the magnitude and (b) the phase of the complex data.

To evaluate the phase shift from the buried gold and $Fe_3O_4$ layers several considerations were applied: Nonresonant SFG can happen at the $CaF_2$-$Fe_3O_4$ as well as the $Fe_3O_4$-$D_2O$ interface for the semi-transparent $Fe_3O_4$ layer. The magnitude of the Fresnel factor for the latter interface is about 7 times larger and thus we consider that it predominantly contributes to the net nonresonant SFG and correct the data with the Fresnel factor of the $Fe_3O_4$-$D_2O$ interface. Similarly, the local oscillator is reflected both at the $CaF_2$-$Fe_3O_4$ and the $Fe_3O_4$-$D_2O$ interface, as depicted in Figure S7(c). Although the latter interface reflects about 35% more light, the resulting interference signal is modulated by the reflection from both interfaces. The phase change upon reflection is at 630 nm about -3 rad for the $CaF_2$-$Fe_3O_4$ interface and around 0.1 for $Fe_3O_4$-$D_2O$. Since the local oscillator field reflected at the $CaF_2$-$Fe_3O_4$ interface does not propagate through the $Fe_3O_4$ layer it is shifted further by about 2.5 rad from the reflected light at the $Fe_3O_4$-$D_2O$ interface. The collective phase shift of -0.6 between the two reflections, together with the different local oscillator intensity due to the magnitude of the reflection, amounts only to a shift of the signal of about 0.2 rad, which concerning our experimental uncertainty is identical to only considering the local oscillator reflection from the $Fe_3O_4$-$D_2O$ interface. Altogether, this results in the phase shift depicted in Figure S7(b), where the signals were normalized by the gold layer, which itself has therefore a phase of 0 rad. The $Fe_3O_4$ nonresonant signal is shifted by about -0.4 rad. The gold layer nonresonant SFG was phase shifted by about -1 rad from the quartz nonresonant signal as shown in Figure S7(a). Followingly, the $Fe_3O_4$ nonresonant SFG is shifted by about -1.4 rad from the quartz nonresonant signal. Thus, the nonresonant SFG from the $Fe_3O_4$ layer is considered majorly real valued in this work as it is shifted approximately by $\pi/2$ from the imaginary quartz signal. Therefore, we used a phase of 0 in the fitting analysis.

## 1.9 Experimental results: The impact of the phase uncertainty in phase-resolved SFG spectroscopy

It was previously discussed that our phase-resolved SFG spectroscopy measurements show a slight drift of the phase associated with the decreased nonresonant signal (see Figure S6). The phase stability was estimated to be $\leq$0.2 rad, which was also the uncertainty range in which the phase of different $Fe_3O_4$ layers with respect to a gold layer varied. Therefore we demonstrate the effect of a phase uncertainty of 0.2 rad, i.e., $\pm$0.1 rad, on our measured spectra both in the OH and OD stretch region in Figure S8. By combining the insights from both the OH and OD stretch region, it can be clearly seen that the phase uncertainty does not influence the deduced peak signs.

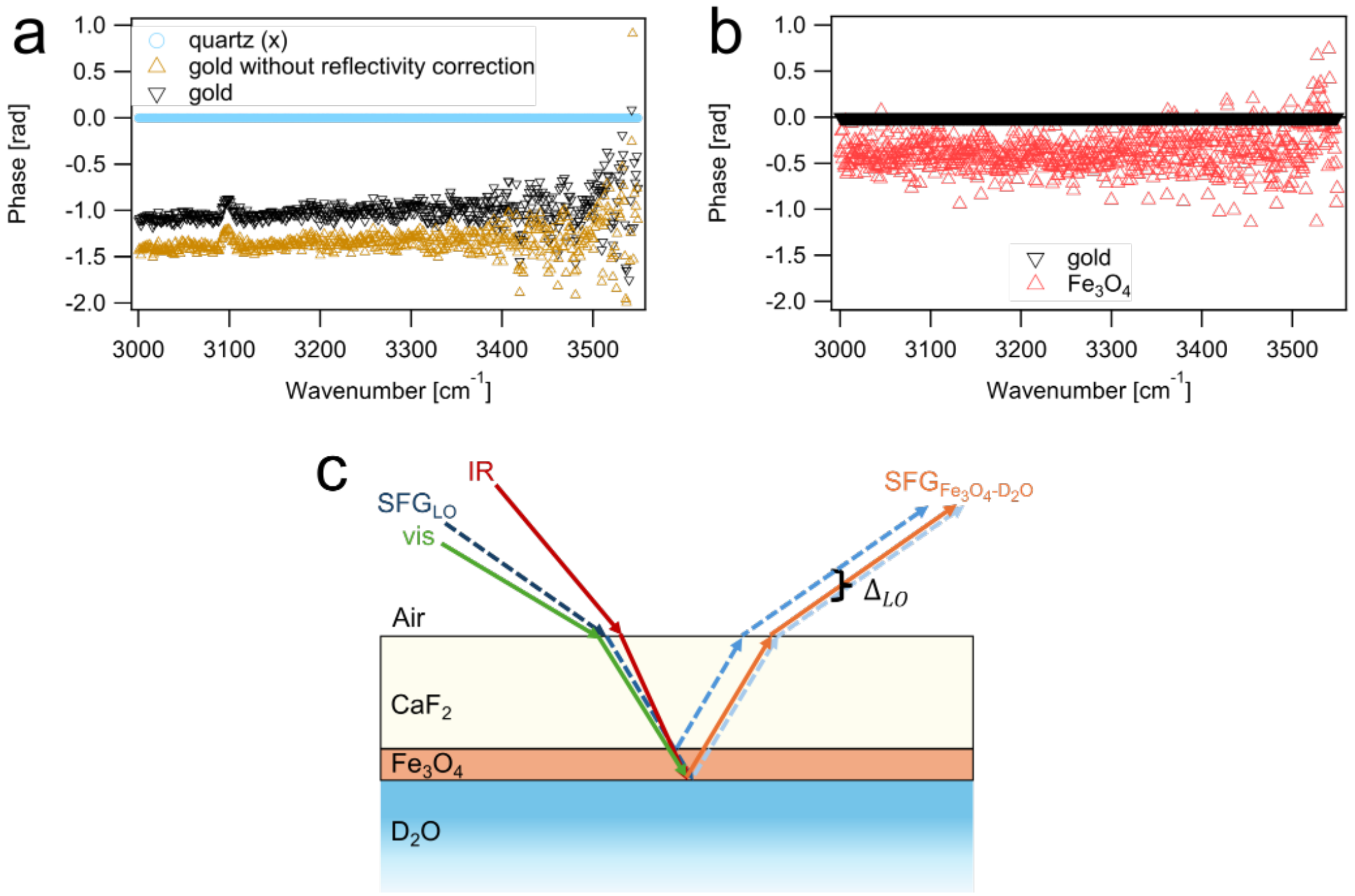

Figure 7: Phase difference from phase-resolved SFG measurements. (a) air-quartz and air-gold interface normalized with the air-quartz spectrum. For comparison the gold data is also shown without the correction of the reflectivity. (b) $CaF_2$-gold-air and $CaF_2$-$Fe_3O_4$-$D_2O$ interface normalized with the $CaF_2$-gold-air spectrum. (c) Sketch of selected beam paths through the sample for the visible (vis), infrared (IR), local oscillator ($SFG_{LO}$) and interfacial SFG ($SFG_{Fe_3O_4}$). $\Delta_{LO}$ indicates the phase shift of the different local oscillator reflections.

## 1.10 Experimental results: SFG spectroscopy comparison of epitaxial grown and sputtered $Fe_3O_4$ layers

The measurements presented in the main text were performed on sputtered $Fe_3O_4$ layers on $CaF_2$ substrates. As XRD measurements were not successful on these layer we could only estimate the crystallinity of the layers by Raman microscopy. As such, we compared the sputtered layers with an epitaxial grown crystalline $Fe_3O_4$(001) layer on MgO, which are characterized in literature (Wangoh et al., 2020). Conventional SFG spectra of the sputtered and epitaxial layers in contact with water at pH 7 are shown in Figure S9(a). The spectra are normalised with the signal of the respective $Fe_3O_4$-$D_2O$ interface. The spectra, and thus the water structure, of both $Fe_3O_4$ layers are qualitatively identical, just the overall intensity measured at the epitaxial layer was a bit higher. The equivalence of the water structure at both layers is further confirmed by the fitted resonant response $\chi_R^{(2)}$, depicted in Figure S9(b).

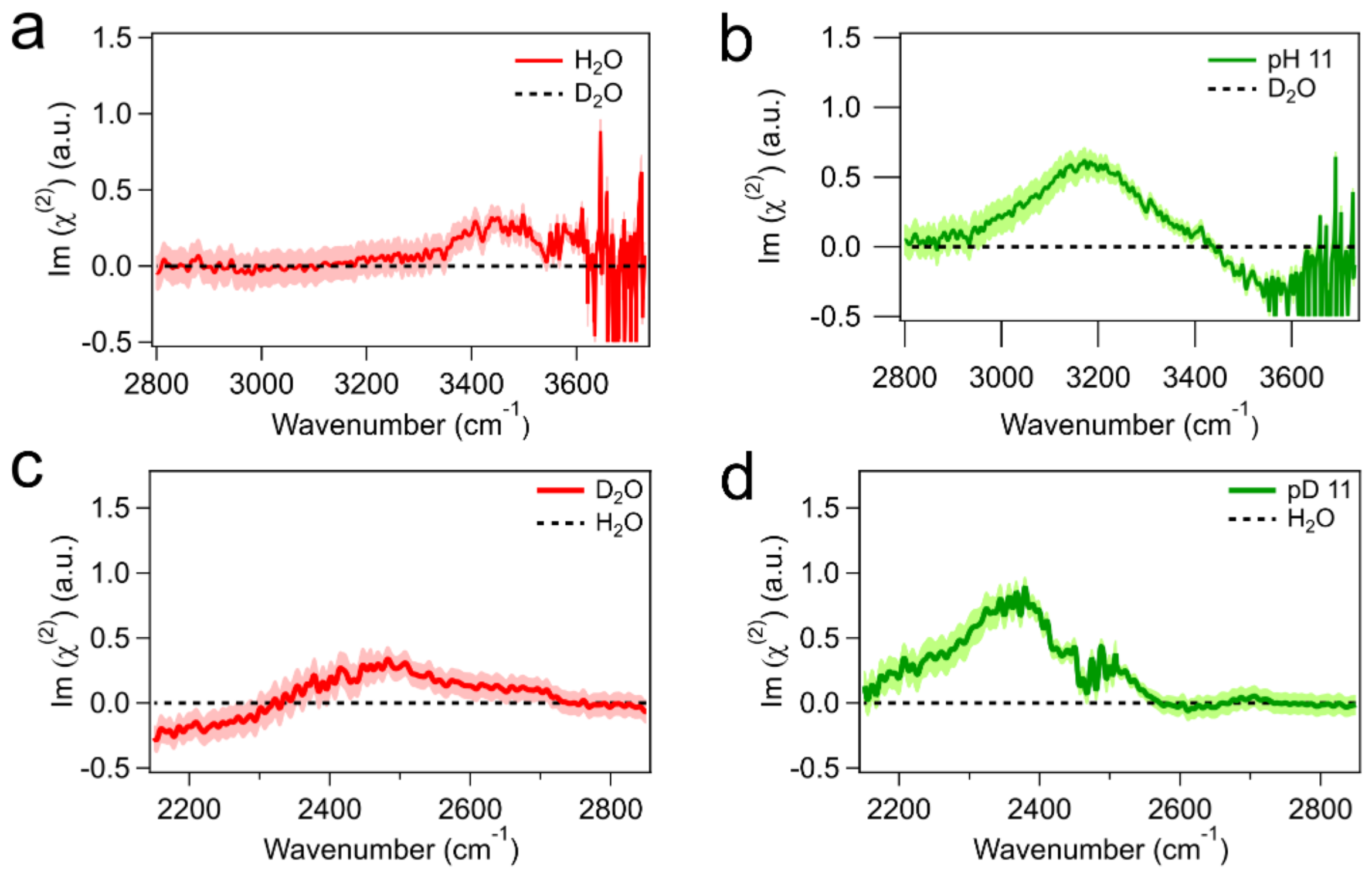

Figure 8: Phase-resolved SFG spectra of $Fe_3O_4$ on $CaF_2$. The spectra are normalised with the $Fe_3O_4$-$D_2O$ interface for the OH stretch region and the $Fe_3O_4$-$H_2O$ interface for the OD stretch region. The shaded area around the data points corresponds to a phase uncertainty of 0.2 rad (±0.1 rad). The $Fe_3O_4$-$H_2O$ interface in the OH stretch region is shown in (a) for neutral pH and (b) for pH 11. The $Fe_3O_4$-$D_2O$ interface in the OD stretch region is shown in (c) for neutral pH and (d) for pD 11.

## 1.11 Experimental results: SFG spectroscopy - surface charge screening experiments

A surface charge can modulate the obtained SFG response from a solid-liquid interface giving rise to third order ($\chi^{(3)}$) contributions, which is already extensively demonstrated in literature (Ong et al., 1992; Rehl et al., 2019; Darlington et al., 2017). Salt screening experiments can be used to gauge beyond surface contributions from a charged interface to the spectrum (Ong et al., 1992). Therefore, we performed conventional SFG spectroscopy measurements of the $Fe_3O_4$-$H_2O$ interface where the ionic strength of the aqueous phase was varied using NaCl up to 100 mM. The spectra from the $Fe_3O_4$-$H_2O$ interface at pH 7 in Figure S10(a) and the $\chi_R^{(2)}$, extracted from fitting, in Figure S10(b) show that there is no dependence of the signal on the ionic strength. Thus, there is no or negligible influence from a surface charge (and therefore $\chi^{(3)}$ contribution) on our spectra, which is fully in accordance with literature where the isoelectric point of the $Fe_3O_4$-$H_2O$ interface was found around pH 6.5-7 (Erdemoğlu and Sarıkaya, 2006; Vereda et al., 2015).

Charge screening experiments performed with pH 11 aqueous solutions (1 mM NaOH) in contact with the $Fe_3O_4$ surface (Figure S10(c) and (d)) showed the typical decreasing trend of the SFG intensity around 3200 $cm^{-1}$ associated with a diminishing $\chi^{(3)}$ response (Ong et al., 1992). As discussed in the main text, this results from a negative surface charge at $Fe_3O_4$. In conclusion, the negligible $\chi^{(3)}$ contribution of the $Fe_3O_4$-$H_2O$ interface at pH 7, warrants a direct comparison of our data with the simulated spectra.

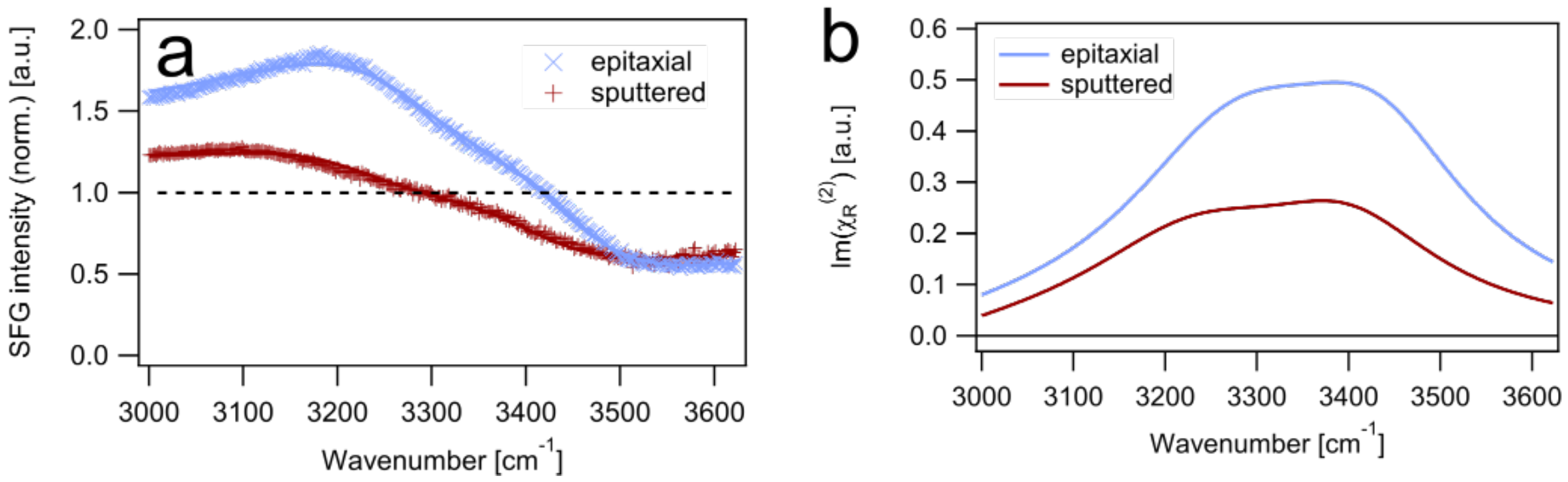

Figure 9: Comparison of Fresnel corrected conventional SFG measurements of epitaxial grown and sputtered $Fe_3O_4$ layers in contact with an aqueous pH 7 solution. (a) Spectra and corresponding fits, (b) plot of the imaginary component from the resonant second order nonlinear susceptibility $\chi_R^{(2)}$ extracted from the fitting analysis.

## 1.12 Experimental results: Layer stability during SFG spectroscopy

The $Fe_3O_4$ layers used within this study were either coated on $CaF_2$ or MgO substrates. It was shown and discussed in the main text that the $Fe_3O_4$ dissolves during measurements, as such it is crucial to ensure that the resulting spectral signatures obtained in SFG spectroscopy only come from the $Fe_3O_4$ layer and not the underlying substrate.

We routinely performed Raman microscopy to investigate the layer intactness after contact with aqueous solutions. Figure S11 compares a $Fe_3O_4$ layer on $CaF_2$ directly after deposition and after extensive contact with aqueous solutions during numerous SFG spectroscopy experiments. The pristine $Fe_3O_4$ layer, shown in (Figure S11(a)) appears homogeneous and also the Raman spectra in Figure S11(c) confirms the presence of $Fe_3O_4$. Several darker ares appear on the layer after usage, depicted in Figure S11(b). However, Raman spectra confirm that these darker area have an identical spectrum to the bright areas and pristine layer. Therefore, also this layer is chemically seen still homogeneously coated with $Fe_3O_4$. The texturing could potentially result from uneven dissolution of the layer or re-precipitation of dissolved products. We further note, that SFG spectra of the $Fe_3O_4$-$H_2O$ interface directly after preparation and after extensive usage are fully comparable.

Another approach to confirm that just the $Fe_3O_4$ layer and not the underlying substrate contributes to our spectra is by comparing our spectra to literature of the $CaF_2$-$H_2O$ (Khatib et al., 2016) and MgO-$H_2O$ (Zelenka and Backus, 2025) interface. According to literature both substrates have a negligible spectral signature, which would be clearly differentiable from our data at the here probed measurement conditions (frequency region and pH range).

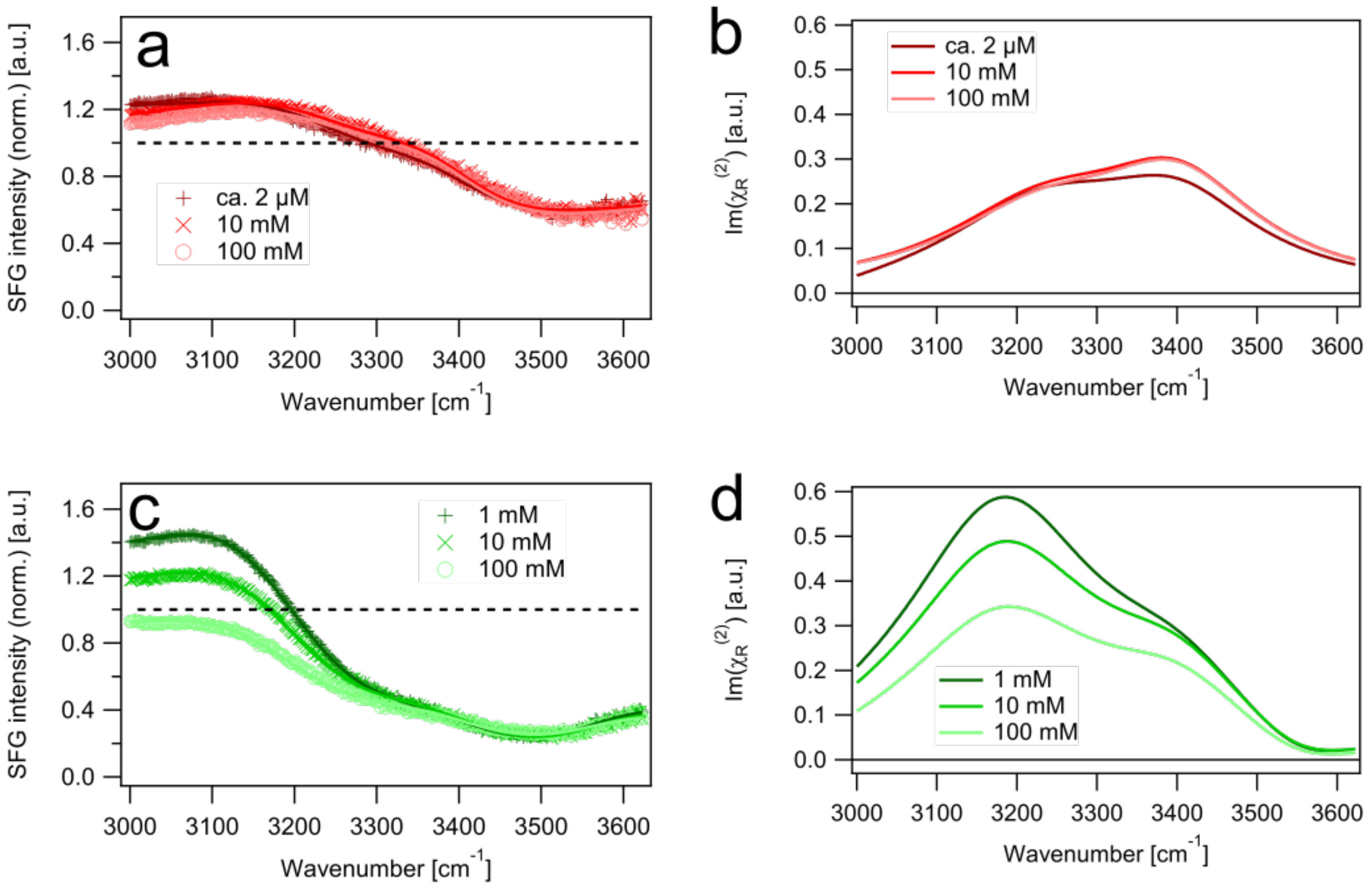

Figure 10: Conventional Fresnel corrected SFG spectra of $Fe_3O_4$ in contact with (a) pH 7 and (c) pH 11 aqueous solutions with additional spectra where NaCl was added to reach 10 and 100 mM ionic strength. Solid lines are fits to the data, the imaginary part of the resonant second order susceptibility $\chi_R^{(2)}$ extracted from the fits is plotted in (b) for pH 7 and (d) for pH 11.

# 2 Theory and Computation

## 2.1 Calculation of the SFG spectra

The resonant SFG spectra were calculated using the correlation-function formalism of Ref. (Khatib and Sulpizi, 2017). For each OH bond, the bond-frame dipole and polarizability derivatives are transformed into the laboratory frame as

$$\dot{\boldsymbol{\mu}}_{i,\text{bond}}^{\text{lab}} \simeq \mathbf{D}_{i,\text{bond}}^{T} \left( \frac{\partial \boldsymbol{\mu}^{\text{bond}}}{\partial r} \dot{r}_{i,\text{bond}} \right), \tag{6}$$

$$\dot{\boldsymbol{\alpha}}_{i,\text{bond}}^{\text{lab}} \simeq \mathbf{D}_{i,\text{bond}}^{T} \left( \frac{\partial \boldsymbol{\alpha}^{\text{bond}}}{\partial r} \dot{r}_{i,\text{bond}} \right) \mathbf{D}_{i,\text{bond}}, \tag{7}$$

where $\mathbf{D}_{i,\text{bond}}$ maps laboratory components into the instantaneous bond frame, whose $z$ axis lies along the OH bond. This single transformation is equivalent to composing the bond-to-molecule and molecule-to-laboratory transformations of the referenced formalism (Khatib and Sulpizi, 2017). These expressions involve two approximations. First, the bond-frame dipole moment and polarizability are Taylor expanded to first order in the OH bond length displacement about a reference geometry, so that their derivatives with respect to $r$ are treated as constant. Second, terms involving the time derivative of $\mathbf{D}_{i,\text{bond}}$ are neglected. This approximation relies on the separation between the high OH stretching motion ($\sim$ 3400 cm$^{-1}$) and the slower molecular reorientation ($\sim$ 1000 cm$^{-1}$) ), whose direct spectral contributions lie predominantly below the OH stretching region.

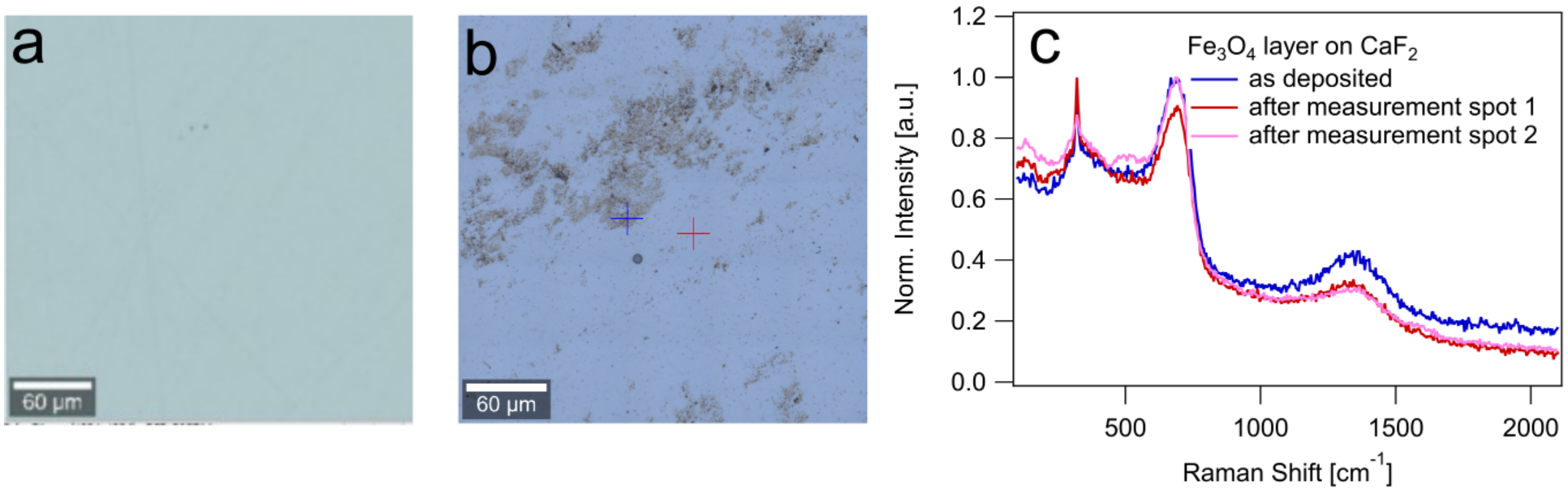

Figure 11: Optical microscopy pictures with 100x magnification of a sputtered $Fe_3O_4$ layer on $CaF_2$ used for the SFG spectroscopy measurements in the main text: (a) measurement directly after deposition, (b) heavily used layer after being in contact with measurement solutions during numerous experiments. The red (spot 1) and blue (spot 2) crosses indicate measurement positions for Raman spectroscopy. (c) A comparison of Raman spectra from the (a) pristine layer and (b) the dark and bright region of the used layer.

With $z$ along the surface normal, the correlation used in the spectral calculation is

$$C(t) = \left\langle \dot{\mu}_z(t) \frac{\dot{\alpha}_{xx}(0) + \dot{\alpha}_{yy}(0)}{2} \right\rangle , \tag{8}$$

giving the resonant response

$$\chi^{(2)}_{\mathrm{SSP},R} = \frac{\chi^{(2)}_{xxz,R} + \chi^{(2)}_{yyz,R}}{2}. \tag{9}$$

The molecular orientations enter through the tensor projections before the two in plane responses are averaged. This average corresponds to the laboratory SSP component with the incidence plane defined as $xz$, under the assumption of an azimuthally isotropic distribution of surface domains. The resulting spectrum shown in Fig. 7 reports $\mathrm{Im}\,\chi^{(2)}_{yyz,R}$.

## 2.2 Dipole moment and polarizability derivatives of OH spieces at the magnetite surface

Table 2: Dipole and polarizability tensor derivatives

| System | $\partial\mu/\partial r\ (z_b)$ [D/Å] | $\partial\alpha_{zz}/\partial r$ [Å$^2$] |
|---|---|---|
| Water | $4.5 \pm 1$ | $4.1 \pm 1$ |
| Water on dissociation | $7.3 \pm 1$ | $7.9 \pm 1.5$ |
| Dissociated OH | $8.4 \pm 1.5$ | $6.2 \pm 1$ |
| Remaining OH | $1.5 \pm 0.5$ | $6.1 \pm 3$ |

Table 3: Derivatives of dipole moment and polarizability with respect to the reaction coordinate $r$ for different OH configurations.

Computing the SFG spectra from simulations requires the dipole moment and polarizability derivatives of the principal OH species at the magnetite surface (Khatib and Sulpizi, 2017). These derivatives were evaluated for the dissociated OH, the remaining OH, and the water on dissociation, as well as for bulk water, using VASP (Kresse and Furthmüller, 1996).

The calculations employed collinear spin-polarized DFT with the optB88-vdW functional and the Dudarev DFT+$U$ correction on Fe 3$d$ orbitals, using $U = 4.50$ eV and $J = 0.89$ eV ($U_{\mathrm{eff}} = 3.61$ eV). The magnetic initialization followed the ferrimagnetic SCV configuration used in the NNP training calculations (Romano et al., 2025), with antiparallel initial moments on octahedral and tetrahedral Fe sites. The spin density was subsequently relaxed self-consistently, without constraining the local magnetic moments. The converged magnetic configurations remained close to the initial ferrimagnetic arrangement. The calculated derivatives therefore incorporate the influence of the converged spin-polarized electronic structure. However, their sensitivity to alternative magnetic configurations or local $Fe^{2+}/Fe^{3+}$ charge-localization patterns was not systematically assessed.

The dipole-moment derivatives were obtained from the hydrogen Born effective charge tensors calculated using density functional perturbation theory, contracted with the unit vector along the corresponding OH bond. The polarizability derivatives were evaluated using a finite difference approximation based on calculations at hydrogen displacements $\Delta r = 0, \pm 0.01, \pm 0.02$ Å along the OH bond, with all other atomic positions and the cell fixed. The polarizability tensors were obtained from the macroscopic electronic dielectric tensors, including local-field effects, using $\boldsymbol{\alpha} = V(\boldsymbol{\varepsilon} - \mathbf{I})/(4\pi)$ in the polarizability-volume convention, where $V$ is the simulation-cell volume and $\mathbf{I}$ is the identity tensor.

The results, averaged over six independent initial structures, are reported in Table 2. All derivatives are expressed in the molecular coordinate frame, with the $z$ axis aligned along the OH bond. Comparing the results for the water OH to the literature (Khatib and Sulpizi, 2017), our derivatives are circa a factor of two greater. The reason of this discrepancy is probably due to the different DFT functionals adopted (Romano et al., 2025). As could have been guessed from the chemical environment and the OH-H bonding network, the dissociated OH and the OH of the water on dissociation share similar values of both the polarization and dipole moment derivatives; interestingly, quite higher than those in the bulk. Opposite behavior is found in the dipole derivative of the reaming OH. The other components of the polarizability tensor derivative have been set to zero, because they were small in absolute value and with relative uncertainty.

## 2.3 Effects of short range interactions on the angular distribution of dipole

The neural network potential (NNP) employed in this work, developed in Ref. (Romano et al., 2025), is a strictly local potential. Short-range interactions are described with high accuracy, while energetic contributions beyond the adopted cutoff of 6 Å are neglected. More precisely, within the Behler–Parrinello framework, the total energy is expressed as a sum of local atomic energies, where each atomic contribution depends on the local atomic environment within a finite cutoff radius, the resulting forces are obtained from its gradient (Behler and Parrinello, 2007). As a consequence, interactions beyond approximately twice the cutoff distance are effectively absent.

In this work, we directly observe the consequences of neglecting long-range electrostatic interactions in the context of dipole–dipole correlations. We therefore take this opportunity to share our findings, with the aim of highlighting potential pitfalls associated with molecular dynamics simulations driven by local potentials, in particular short range ML potentials, intensively employed in recent years.

We focus on the orientational ordering of water molecules at the magnetite–water interface. To this end, we computed the probability density of $\cos(\theta)$ where $\theta$ is the angle between the water dipole moment and the surface normal, as explained in the main text. In Fig. 12a we show this distribution resolved by individual layers parallel to the interface with the first two layers identified by the visible minima of the oxygen density distribution (3.8 Å and 7.2 Å) (see (Romano et al., 2025)) while the remaining water slab subdivided into equally spaced layers. Strong interactions with the magnetite surface induce a pronounced orientational anisotropy in the first water layer (within 3.8 Å from the surface), characterized by three distinct peaks. The positive peak corresponds to water molecules adsorbed on octahedral iron rows and pointing away from the surface, while the two negative peaks arise from the water on dissociation and water molecules above tetrahedral iron rows. The second water layer (up to 7.2 Å) exhibits a much weaker orientational preference, forming a complementary profile to that of the first layer. Beyond this distance, water molecules start to no longer be directly influenced by the surface and interact predominantly with the surrounding, already polarized, interfacial water. The physics captured by the short range potential at this level is the local part of the dipole–dipole interaction, which energetically favors parallel alignment of neighboring dipoles. However, such dipolar ordering is intrinsically accompanied by charge separation (proportional to the total dipole moment projected into the surface normal) that generates a macroscopic electric field. This "depolarization field" would give rise to a free-energy penalty proportional to the square of the polarization and would enforce dielectric screening by the opposite arrangement of the dipoles of the subsequent layers. This collective electrostatic response is entirely absent in short-range interaction potentials, as previously reported in studies observing the same phenomenon (Gao and Remsing, 2022; Cox, 2020; Niblett et al., 2021). Within the present framework, the only mechanism opposing complete polarization of the water slab is therefore entropic. Consistently, from the third water layer onward we observe a linearly sloped distribution of $\cos(\theta)$, indicating a net average orientation of water molecules with their hydrogen atoms pointing toward the surface.

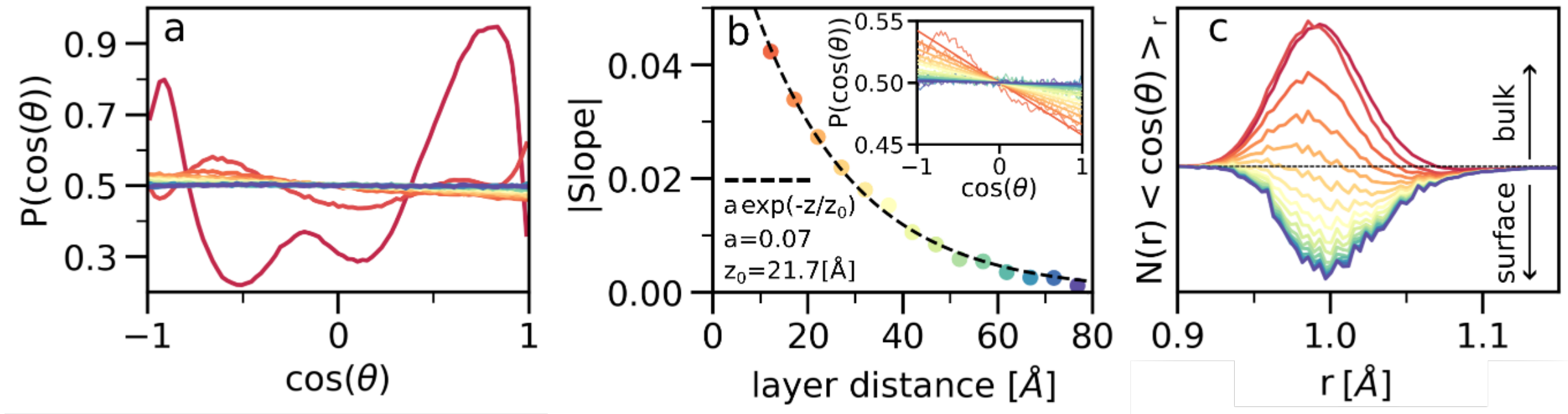

Figure 12: Short range potentials allows unphysical long range orientational anisotropy of water at the magnetite interface. Results for different water layers are colored from red (surface) to purple (80 Å less than half water column). a) Probability density of $\cos(\theta)$ for the water dipoles in different layers. b) Slope of linear fits to the probability densities of $\cos(\theta)$ (from the third layer on) as function of the distance of the layer from the surface. The dashed line is an exponential fit to the slopes with a decay length of $z_0 = 21.7$Å. The probability densities of $\cos(\theta)$ from the third layer on, together with the corresponding linear fits, are shown in the inset with a magnified y-axis. c) Orientational average $\langle\cos(\theta)\rangle_r$ for OH bonds of a given length $r$ times the number $N(r)$ of OH bonds with that length. Each curve corresponds to a cumulative average up to the layer indicated by the color. The dashed line indicates the zero.

To assess the spatial extent over which the exclusion of long-range interactions affects the orientational distribution, three systems with increasing water thickness (55, 85, and 160 Å) were simulated, containing 3152, 6274, and 12048 water molecules and spanning approximately 29 ns, 29 ns, and 14 ns of simulation time, respectively (the full trajectory of the largest system is available on Zenodo, DOI: 10.5281/zenodo.15658131). A finite-size effect is observed in the orientational anisotropy, quantified as the slope of the linear fit, for the two thinner water slabs, which display a linear recovery of the bulk-like isotropic orientational distribution. In these cases, the recovery is solely enforced by the periodic boundary conditions. This linear trend is consistent with previous reports in the literature (Gao and Remsing, 2022) and with the supporting information of Ref. (Litman et al., 2023). Only in the largest system does the recovery of the bulk-like, isotropic orientational distribution follow an exponential decay. The associated orientational anisotropy length scale at 300 K is approximately 22 Å (Fig. 12b) We emphasize that this extended polarization profile is not expected to persist in simulations employing a proper long-range electrostatic description. The observed exponential decay arises here exclusively from the propagation of short-range orientational correlations, as the NNP includes a force cutoff at approximately 12 Å. This behavior can be interpreted as a decorrelation process of water dipole orientations away from a preferentially aligned interfacial layer, analogous to an Ising-like system with nearest-neighbor interactions. In contrast, in the presence of full Coulomb interactions, collective dipolar alignment would generate a macroscopic depolarization field, leading to a free-energy penalty proportional to the square of the polarization. This energetic cost enforces efficient dielectric screening, resulting in a rapid decay of orientational correlations and a strong suppression of net polarization beyond the first few interfacial layers. The persistence of long-ranged dipolar ordering observed here should therefore be understood as an artifact of the local nature of the interaction potential, rather than as a physical property of interfacial water. Fig. 12c quantitatively shows the OH bond orientation average for a given bond length of cumulative water layers at the interface. As isotropic water layer would add zero contribution to this average. the long range polarization of water can be clearly evinced from the continuous reshaping of the curves observed even including farther water layer.

## 2.4 The hydroxylated surface model

The SFG experiments were performed at pH 7 and pH 11. While pH is not a direct control parameter in classical molecular dynamics simulations, the experimental conditions suggest that the magnetite surface becomes increasingly deprotonated at high pH, resulting in a larger abundance of surface hydroxyl groups. The aim of this section is therefore not to provide a quantitative or electrochemically complete description of the pH dependence, but rather to obtain qualitative insight into how an increased degree of surface deprotonation may affect the local interfacial structure.

To this end, we constructed a simplified hydroxylated surface model by removing one hydrogen atom per surface unit cell from the interfacial water layer, thereby mimicking a more deprotonated surface termination. We emphasize that this model should be regarded as a schematic proxy for a high-pH surface state, rather than a full representation of the pH 11 electrochemical interface. In particular, counterions and long-range electrostatic effects are not included, and the model is intended to probe only local structural consequences of surface deprotonation.

Figure 13 compares the hydrogen density profiles in the first water layer for the neutral and hydroxylated surface models. A clear reduction of the first tetrahedral peak, accompanied by an increase of the second tetrahedral peak, is observed for the hydroxylated surface. This behavior indicates that the reduced availability of hydrogen atoms at the surface hinders the dissociation of interfacial water molecules. Quantitatively, the number of dissociated water species decreases by approximately 30%, as estimated from the reduction of the area of the first (dissociated) peak. In the absence of dissociation sites containing dissociated hydrogens, a larger fraction of water molecules adopts configurations in which the molecular dipole is oriented toward the surface and both hydrogen atoms participate in hydrogen bonding with surface oxygen atoms (corresponding to scenario (b) in Fig. 4 of the main text). These configurations give rise to the enhanced second tetrahedral peak, which increases by a comparable fraction.

As a direct consequence of the construction of the model surface, the hydroxylated system contains a higher density of surface OH groups than the neutral case. This modification also affects the octahedral hydrogen density peak, where the shoulder on the high-distance side is reduced. This feature originates from OH bonds of water molecules adsorbed on octahedral iron sites, whose population is diminished in the hydroxylated surface model.

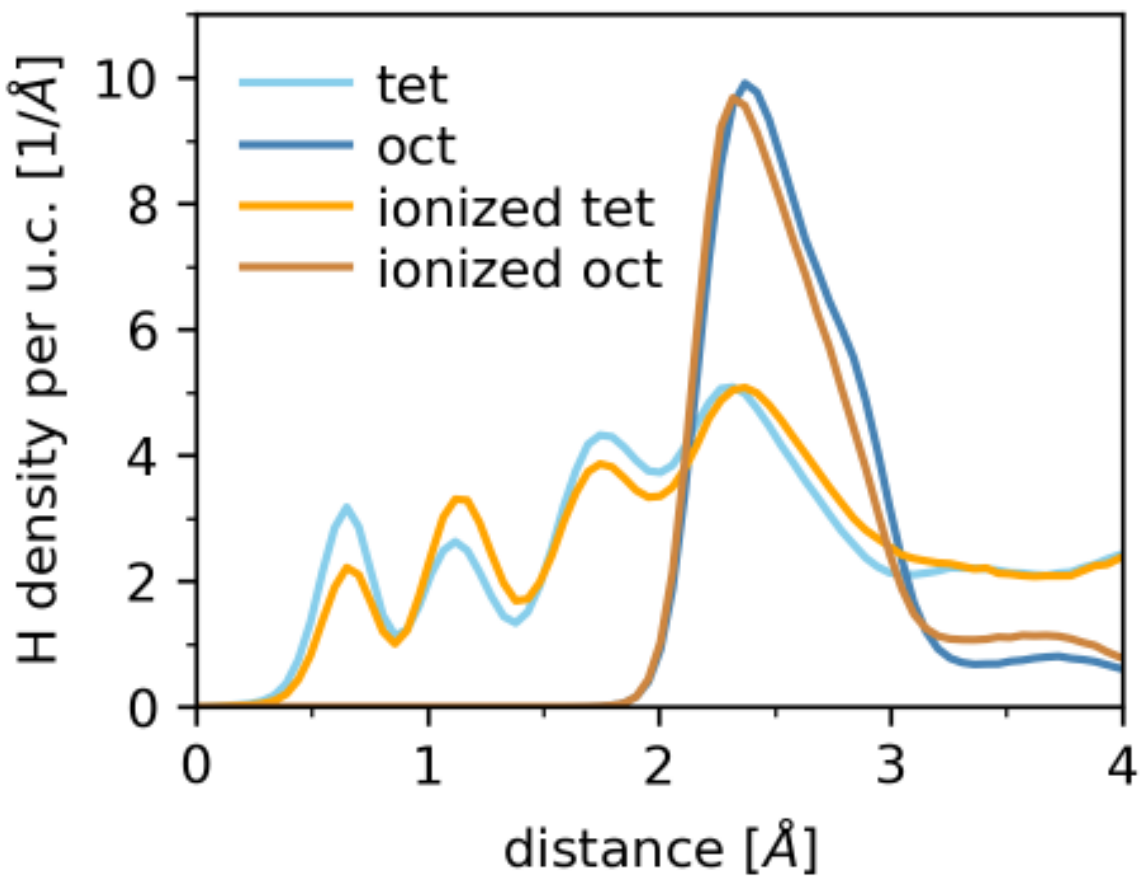

Figure 13: Density profile of the hydrogen atoms on the magnetite surface for the neutral pH water (blues) and the hydroxylated water (yellows). The curves distinguish both between the location at the surface: on the octahedral iron row (dark) or on the tetrahedral one (light)